\documentclass[manuscript]{aastex}
\usepackage{natbib}
\def\intl{\textit{INTEGRAL}}
\def\rxte{\textit{RXTE}}

\def\spitzer{\textit{Spitzer}}

\def\chis2{$\chi^2$}
\def\msun{$M_{\odot}$}
\def\rsun{$R_{\odot}$}

\def\wm2{W~m$^{-2}$}
\def\mic{$\mu$m}
\def\cm2{cm$^2$}
\def\se1{s$^{-1}$}

\def\grs1915{GRS~1915$+$105}

\shorttitle{Multi-wavelength studies of \grs1915}
\shortauthors{Rahoui et al.}

\begin{document}


\title{Long-term multi-wavelength studies of \grs1915\\
I. A high-energy and mid-infrared focus with \rxte/\intl\ and \spitzer}


\author{F. Rahoui\altaffilmark{1,2,3}}
\email{frahoui@cfa.harvard.edu}
\author{S. Chaty\altaffilmark{1}}
\author{J. Rodriguez\altaffilmark{1}}
\author{Y. Fuchs\altaffilmark{1}}
\author{I.F. Mirabel\altaffilmark{1,4}}
\and
\author{G.G Pooley\altaffilmark{5}}




\altaffiltext{1}{Laboratoire AIM, CEA/DSM - CNRS - Universit\'e Paris Diderot,
  Irfu/Service d'Astrophysique, B\^at. 709, CEA/Saclay, F-91191 Gif-sur-Yvette, France}
\altaffiltext{2}{AstroParticule \& Cosmologie (APC) / Universit\'e Paris VII / CNRS / CEA / 
Observatoire de Paris, B\^at. Condorcet, 10 rue Alice Domon et L\'eonie Duquet, 75205 Paris Cedex 13, France}
\altaffiltext{3}{Harvard-Smithsonian Center for Astrophysics, 60 Garden Street, Cambridge, MA, 02138, USA}
\altaffiltext{4}{Instituto de Astronomia y Fisica del Espacio-CONICET. cc 67, suc. 28 (C1428ZAA). Buenos Aires - Argentina}
\altaffiltext{5}{ Astrophysics, Cavendish Laboratory, J. J. Thomson Avenue, Cambridge CB3 0HE, UK}


\begin{abstract}
To date, mid-infrared properties of Galactic black hole binaries have barely been investigated in the framework of multi-wavelength campaigns. Yet, studies in this spectral domain are crucial to get complementary information on the presence of dust and/or on the physical processes such as dust heating and thermal \textit{bremsstrahlung}. Here, we report a long-term multi-wavelength study of the microquasar \grs1915. On the one hand, we aimed at understanding the origins of the mid-infrared emission, and on the other hand, at searching for correlation with the high-energy and/or radio activities. We observed the source at several epochs between 2004 and 2006 with the photometer IRAC and spectrometer IRS, both mounted on the \spitzer\ \textit{Space Telescope}. When available, we completed our set of data with quasi-simultaneous \rxte/\intl\ high-energy and/or Ryle radio observations from public archives. We then studied the mid-infrared environment and activities of \grs1915\ through spectral analysis and broad band fitting of its radio to X-ray spectral energy distributions.
We detected polycyclic aromatic hydrocarbon molecules in all but one IRS spectra of \grs1915\ which unambiguously proves the presence of a dust component, likely photoionised by the high-energy emission. We also argue that this dust is distributed in a disc-like structure heated by the companion star, as observed in some Herbig Ae/Be and isolated cool giant stars. Moreover, we show that some of the soft X-ray emission emanating from the inner regions of the accretion disc is reprocessed and thermalised in the outer part. This leads to a mid-infrared excess that is very likely correlated to the soft X-ray emission. We exclude thermal \textit{bremsstrahlung} as contributing significantly in this spectral domain.
\end{abstract}


\section{Introduction}

In a multi-wavelength study of microquasars, the infrared presents a particular interest as the accretion disc, the jets, or the companion star may all be detected. Nevertheless, most of the previous studies focused on the accretion$-$ejection phenomena as seen is X-ray/near-infrared/radio correlations, and did not  take the mid-infrared (MIR) emission into account \citep[see \textit{e.g.}][]{1998Mirabel, 2002Ueda, 2002Corbel, 2003Corbel, 2003Chaty, 2005Homanb, 2006Chatya, 
2006Russell}.  Yet, getting both MIR photometric and spectroscopic information is crucial to investigate the presence of dust, the disc illumination, thermal \textit{bremsstrahlung} from the accretion disc's wind, or the 
contribution of relativistic ejecta. 

\subsection{\grs1915}

Discovered by the WATCH all-sky X-ray monitor on board the \textit{GRANAT} satellite, on 1992 August 15 \citep{1992Castro, 1994Castro}, \grs1915\ is the first microquasar in which apparent superluminal radio ejecta were detected \citep{1994Mirabelb}. The nature of its companion star was the subject of debate until \citet{2001Greinera} unambiguously showed that it was a K/M red giant by detecting CO absorption features in its near-infrared (NIR) spectrum. Moreover, the orbital period of the system and the mass of the compact object were found to be $33.5\pm1.5$~days \citep[recently refined to $30.8\pm0.2$~day,][]{2007Neil} and $14\pm4$~\msun, respectively \citep{2001Greinerb}. The inclination is $66^{\circ}\pm2^{\circ}$, and estimates of the distance fall in the range 6$-$12~kpc \citep{1996Chaty, 1999Fender, 2004Chapuis}.

\grs1915\ is strongly variable, on time scales from seconds to days. Using extensive \rxte\ timing observations, \citet{2000Belloni} showed that its X-ray behaviour could be divided in 12 distinct luminosity classes \citep[up to 14 today,][]{2002Klein, 2003Hannikainen}, and that the source was carrying out transitions between three canonical spectral states, labelled A, B (strong disc domination), and C (corona-dominated, no disc). Previous multi-wavelength studies showed the existence of a strong connection between the accretion disc instabilities and plasma outflows. In particular, discrete ejecta, emitting through optically thin synchrotron, are believed to be triggered during the transition between the C and A states, expanding adiabatically in the environment and detectable gradually from the NIR to the radio domains \citep[see \textit{e.g.}][]{1996Mirabel, 1997Fender, 1998Mirabel, 1998Eikenberry, 2000Eikenberry, 2008Rodrigueza, 2008Rodriguezb}. Moreover, the other known class of radio ejecta $-$ the compact jets, emitting simultaneously from the radio to the NIR through optically thick synchrotron $-$ are only detected in the $\chi$ luminosity class, which is only seen in the C state. The presence of such a jet is characterised by a flat spectrum with a roughly constant flux density between 50 and 100~mJy.  Long periods of the $\chi$ luminosity class during which compact jets are present are called \textit{plateau}, and often precede or follow a giant ejection \citep[see \textit{e.g.}][]{1996Foster, 1997Pooley, 1999Fender, 2000Dhawan, 2002Klein, 2003Fuchsb}. 

\subsection{Previous MIR studies of microquasars}

\citet{2002Koch} presented an \textit{ISO} spectrophotometric study of Cygnus~X$-$3 in quiescence, 
and found the MIR continuum to be due to free-free emission from the winds of the Wolf-Rayet companion star 
with perhaps a contribution from a cold dust component \citep[see][for similar conclusions on SS~433]{2006Fuchs}. Moreover, \citet{2003Fuchsa} reported ISOCAM photometric data of the \grs1915 obtained at two different epochs, during a flaring activity and a \textit{plateau} state. On the one hand, they showed that, despite strong uncertainties, the MIR flux of \grs1915\ had likely increased between the two observations, and they argued that during the \textit{plateau}, the MIR emission of the source was likely due to a compact jet, without excluding \textit{bremsstrahlung}. They, on the other hand, excluded the contribution of dust, which is the opposite of the conclusion reached by \citet{2006Muno} to explain the MIR excess they detected in the \spitzer/IRAC SEDs of A0620$-$00 and XTE~J1118+480 in quiescence. Finally, \citet{2007Migliari} argued that X-ray/UV irradiation of the disc in the thermal state, and compact jet in the hard state might be responsible for the excess they detected in the MIR emission of GRO~J1655$-$40; the same conclusion was reached  by \citet{2007Gallo} concerning the quiescence of A0620$-$00, V404~Cyg, and XTE~J1118+480, claiming that dust component is not statistically necessary. 
\newline

In this paper, we report a long-term multi-wavelength study of \grs1915\, focusing on its spectroscopic and photometric MIR emission.  It aimed at understanding its origins, as well as its possible connection with the high-energy and radio domains. The observations and the data analysis are presented in Sect.~2, while Sect.~3 and Sect.~4 are devoted to the analysis of the broad band X-ray to MIR SEDs of the source built with high-energy and \spitzer\ data. We discuss the outcomes in Sect.~5 and we give our conclusions in Sect.~6.

\section{Observations and data analysis}

We performed, between 2004 October 2 and 2006 June 5, 16 photometric and spectroscopic observations of \grs1915\ with the IRAC photometer 
and the IRS spectrometer (PI Y. Fuchs), both mounted on the \textit{Spitzer Space Telescope}.  
Moreover, we completed our set of MIR observations with quasi-simultaneous high-energy observations, including \intl\ data already presented in great detail in \citet{2008Rodrigueza, 2008Rodriguezb} (revolution \#373), and several monitoring observations with \rxte\ (PI Morgan, the data are immediately public). 
We finally also made use of observations of \grs1915~obtained at 15~GHz with the Ryle Radio Telescope (PI G. G. Pooley). Table~\ref{logobs} lists all the data we used in this study. In the following, for each quasi-simultaneous X-ray/MIR/radio observation, we will refer to the integer part of the MIR observation date. 

\subsection{\textit{Spitzer}'s MIR observations}

\grs1915~was observed with IRAC at 3.59, 4.50, 5.80, 
and 8.00~\mic. Each image was the combination of 24 sub-exposures of 10~s each, giving a total 
integration time of 240~s in each filter.  We performed photometry on the Basic Calibration 
Data (BCD) using the software \textit{MOsaicker and Point source EXtractor} (\textit{MOPEX}) v18.2.2. 
BCD data are raw data on which the \textit{Spitzer} pipeline performs dark subtraction, multiplexer bleed correction, 
detector linearisation, flat fielding, cosmic ray detection, and flux calibration. 
We used \textit{MOPEX} for pointing refinement, mosaicking, coaddition, and fluxes measurement through PSF fitting in a 3-pixel radius aperture. 
They  were then scaled to a 10-pixel aperture using the aperture-correction factors as given in the IRAC manual\footnote{http://ssc.spitzer.caltech.edu/irac/dh/iracdatahandbook3.0.pdf}. The source was always detected, 
in all filters, and the absorbed fluxes are listed in Table~\ref{fluxirac1915}. The uncertainties include the 3\% systematic errors due to flux calibration instabilities \citep{2005Reach}.
\newline

\begin{figure*}
\begin{center}
\begin{tabular}{cc}
MJD~53280&MJD~53299\\
\includegraphics[height=6.5cm,width=4cm, angle=90]{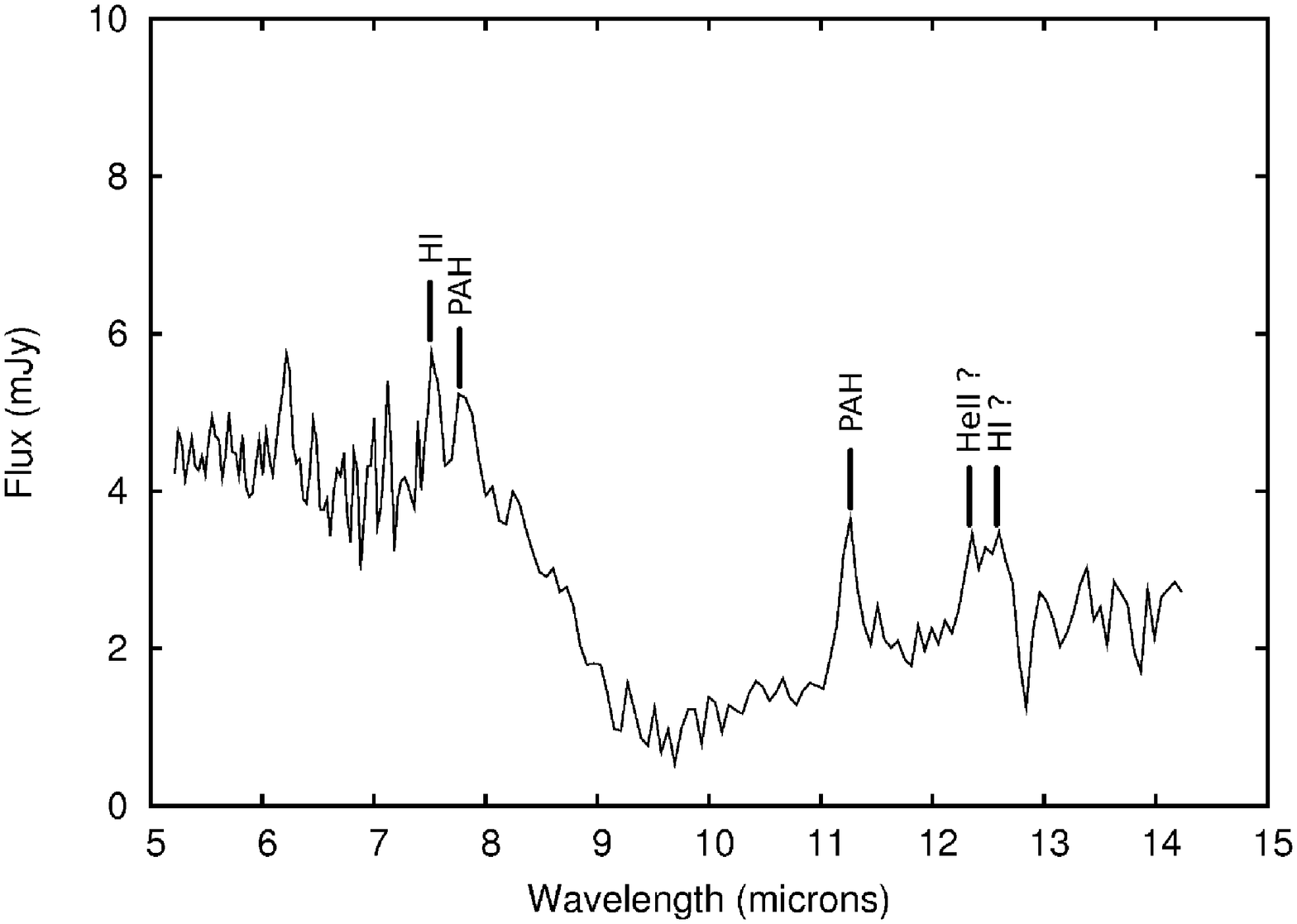}&\includegraphics[height=6.5cm,width=4cm, angle=90]{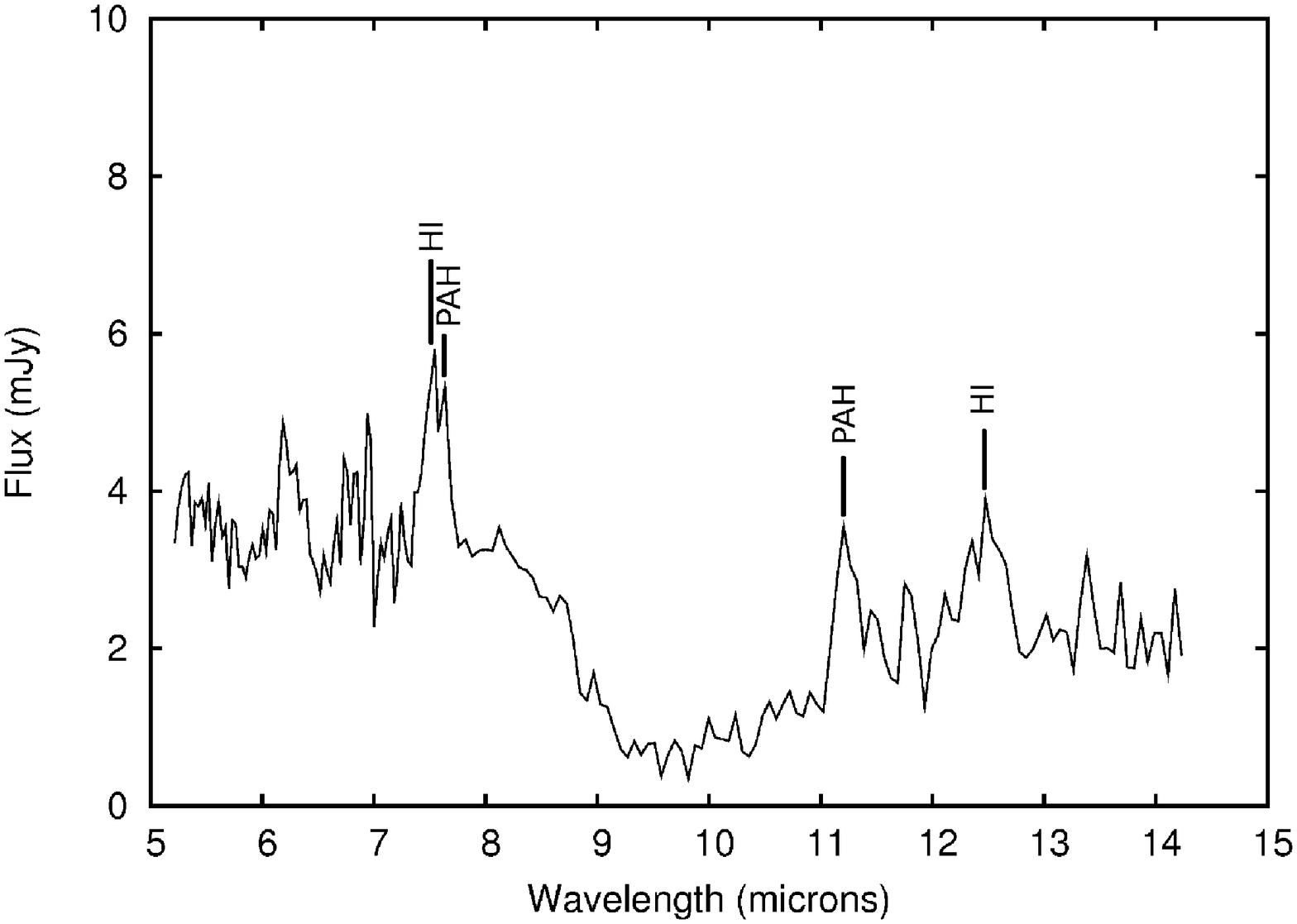}\\
MJD~53484&MJD~53511\\
\includegraphics[height=6.5cm,width=4cm, angle=90]{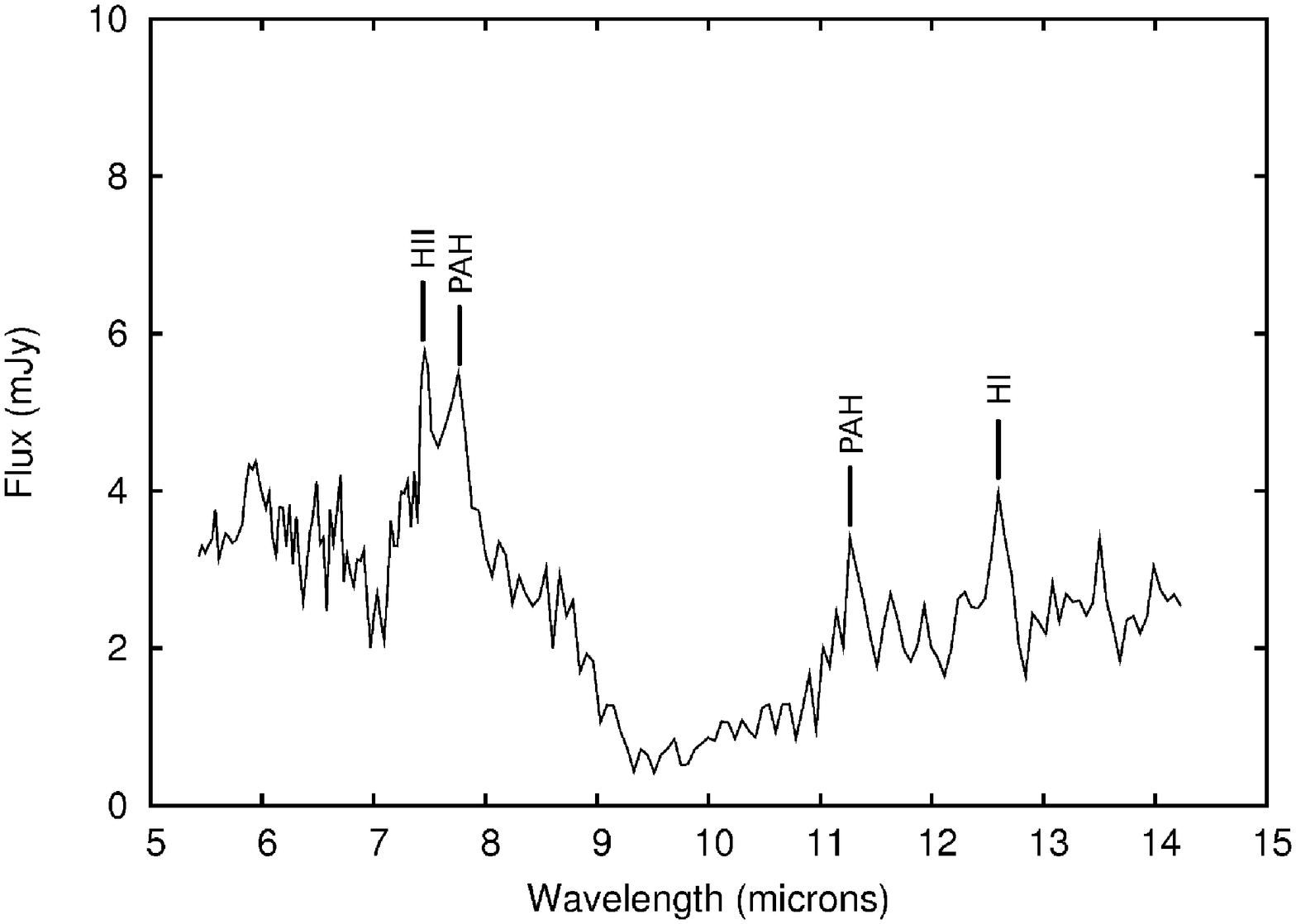}&\includegraphics[height=6.5cm,width=4cm,angle=90]{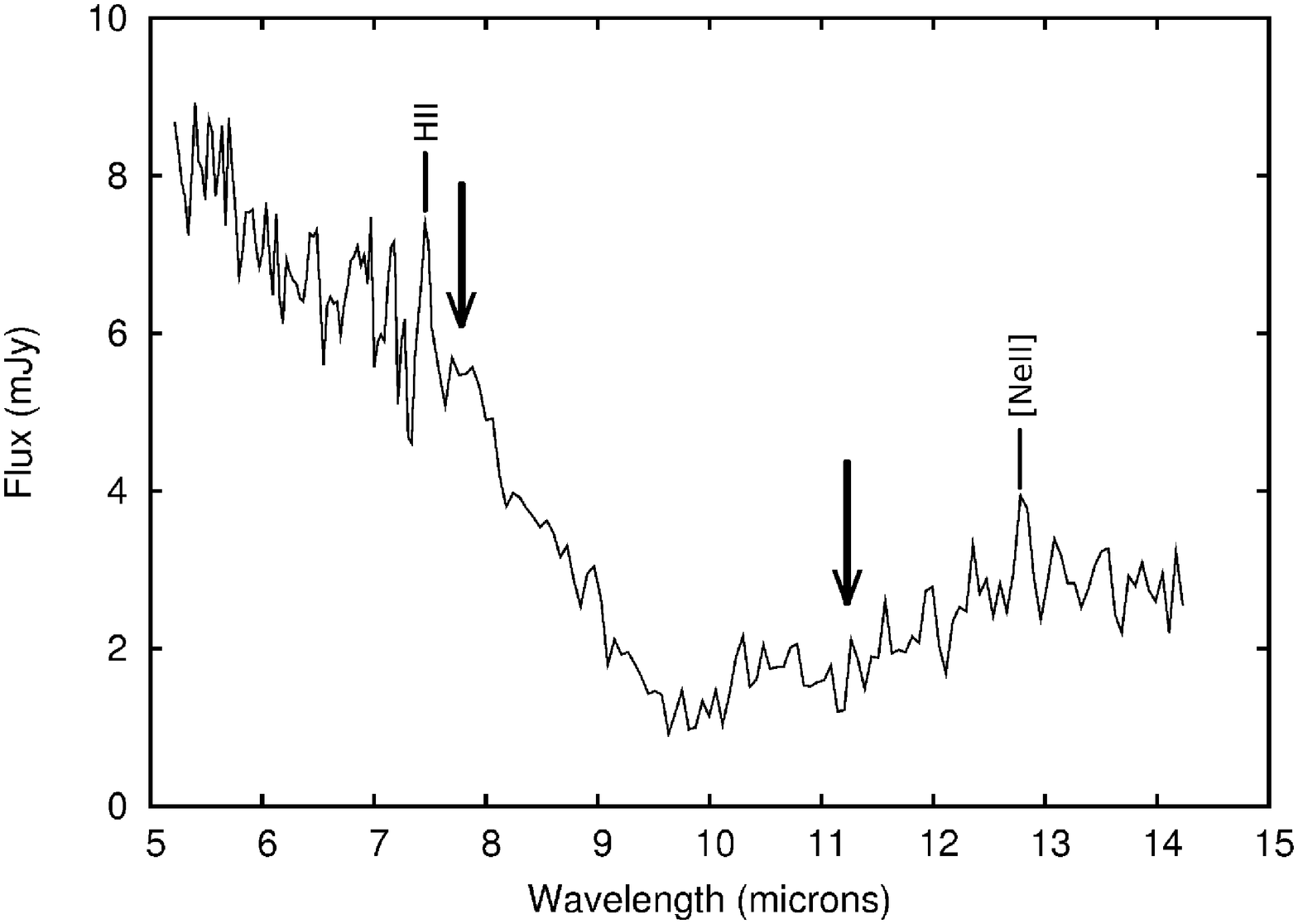}\\
MJD~53660&MJD~53661\\
\includegraphics[height=6.5cm,width=4cm, angle=90]{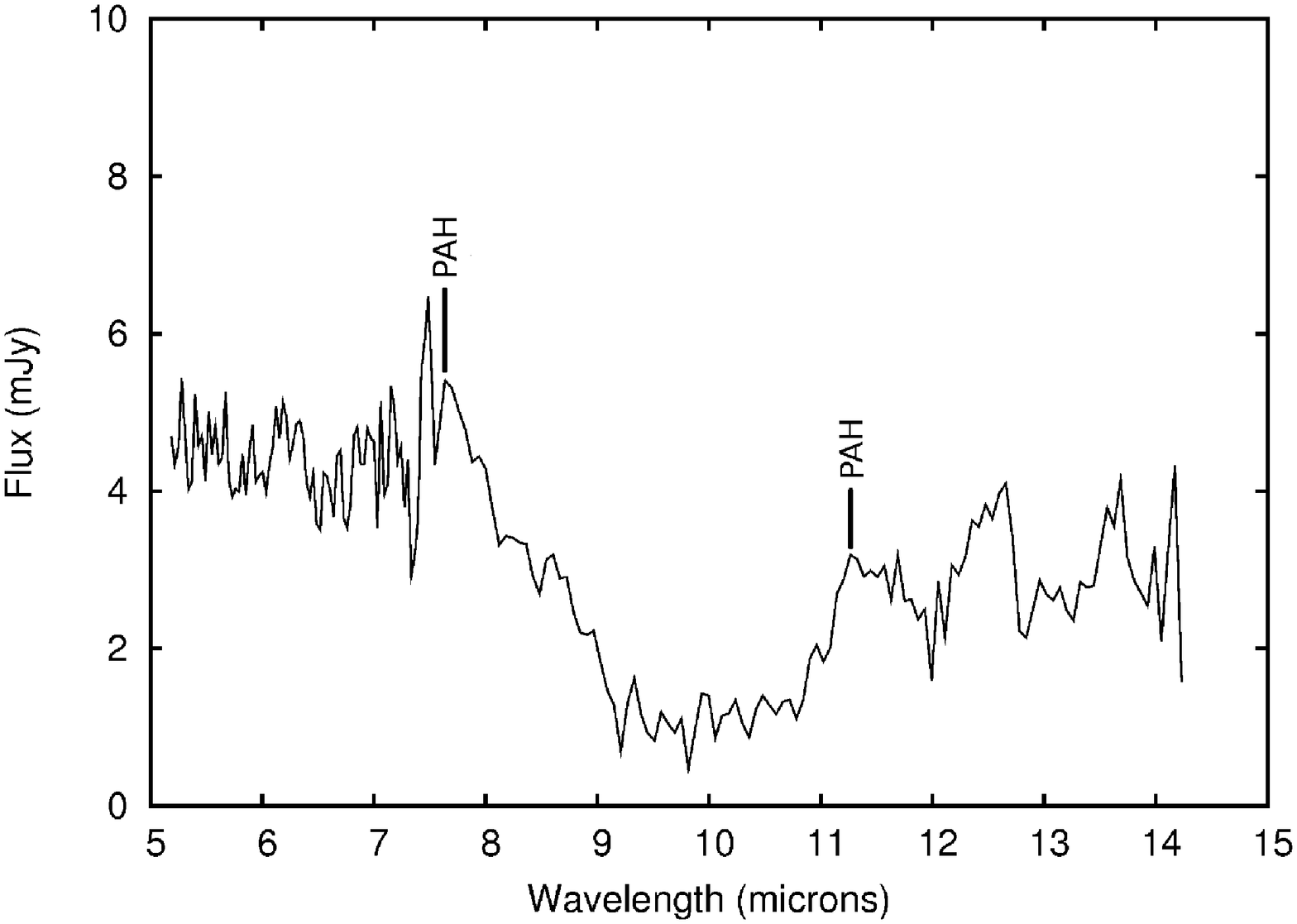}&\includegraphics[height=6.5cm,width=4cm, angle=90]{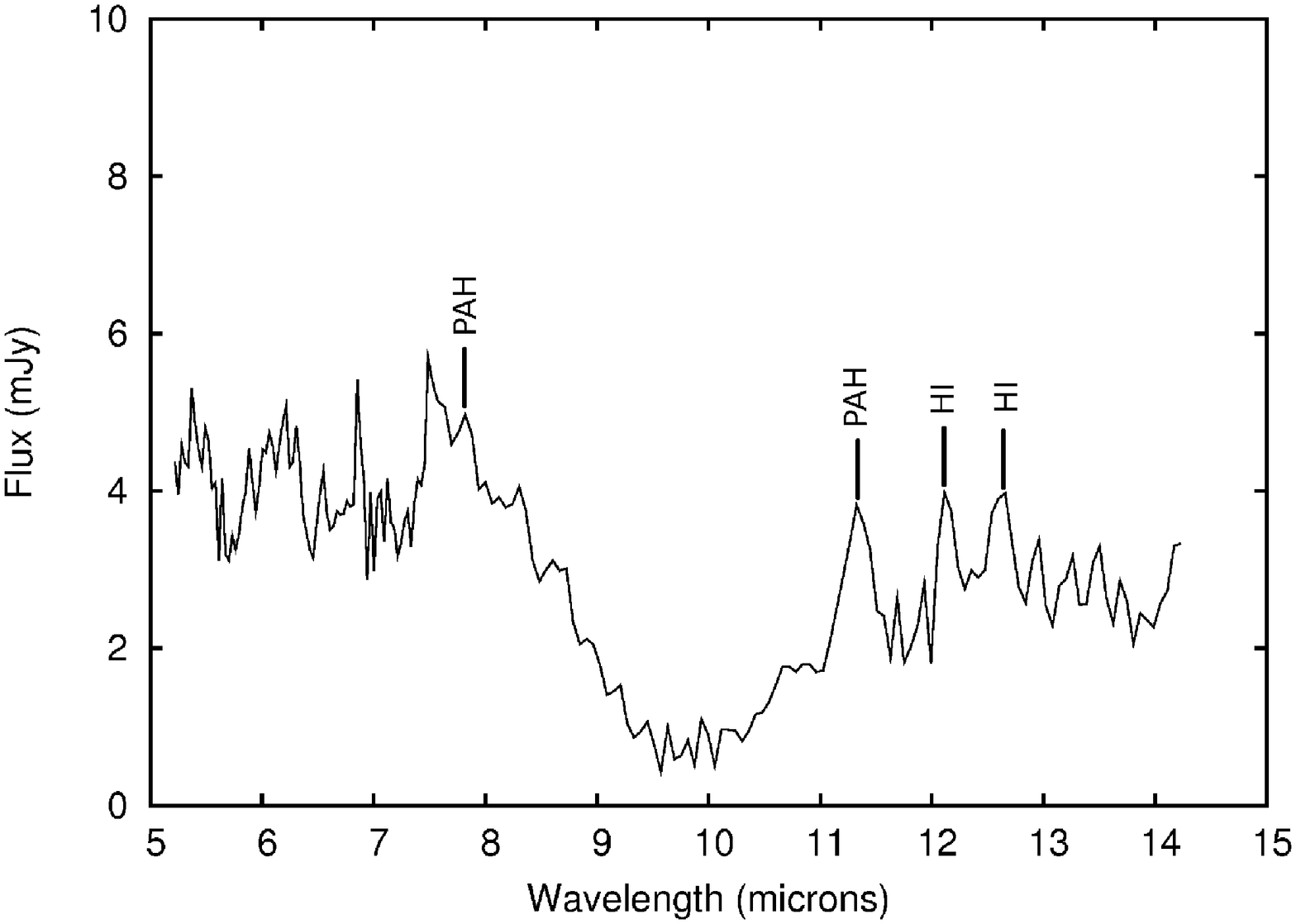}\\
MJD~53851&MJD~53874\\
\includegraphics[height=6.5cm,width=4cm, angle=90]{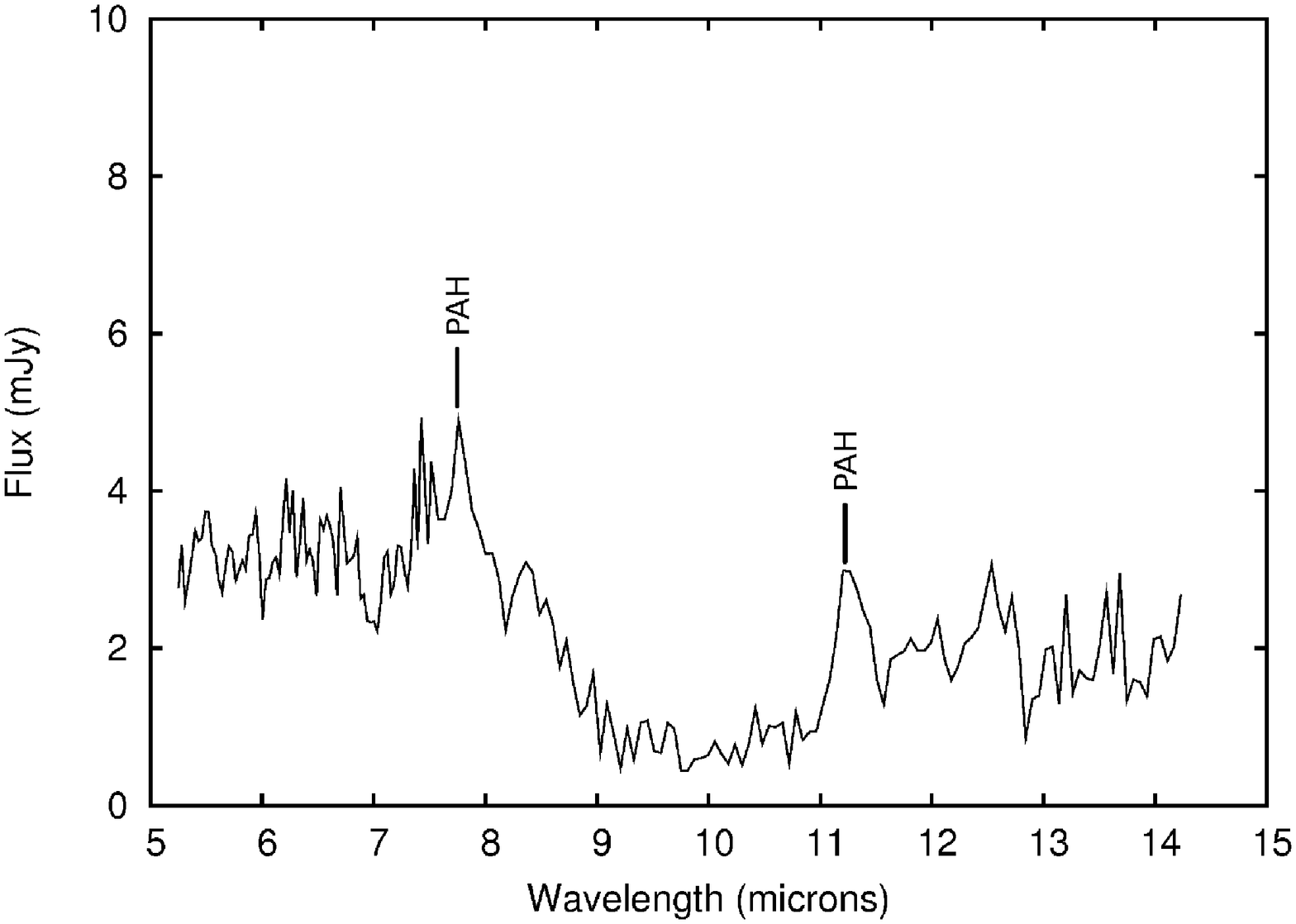}&\includegraphics[height=6.5cm,width=4cm, angle=90]{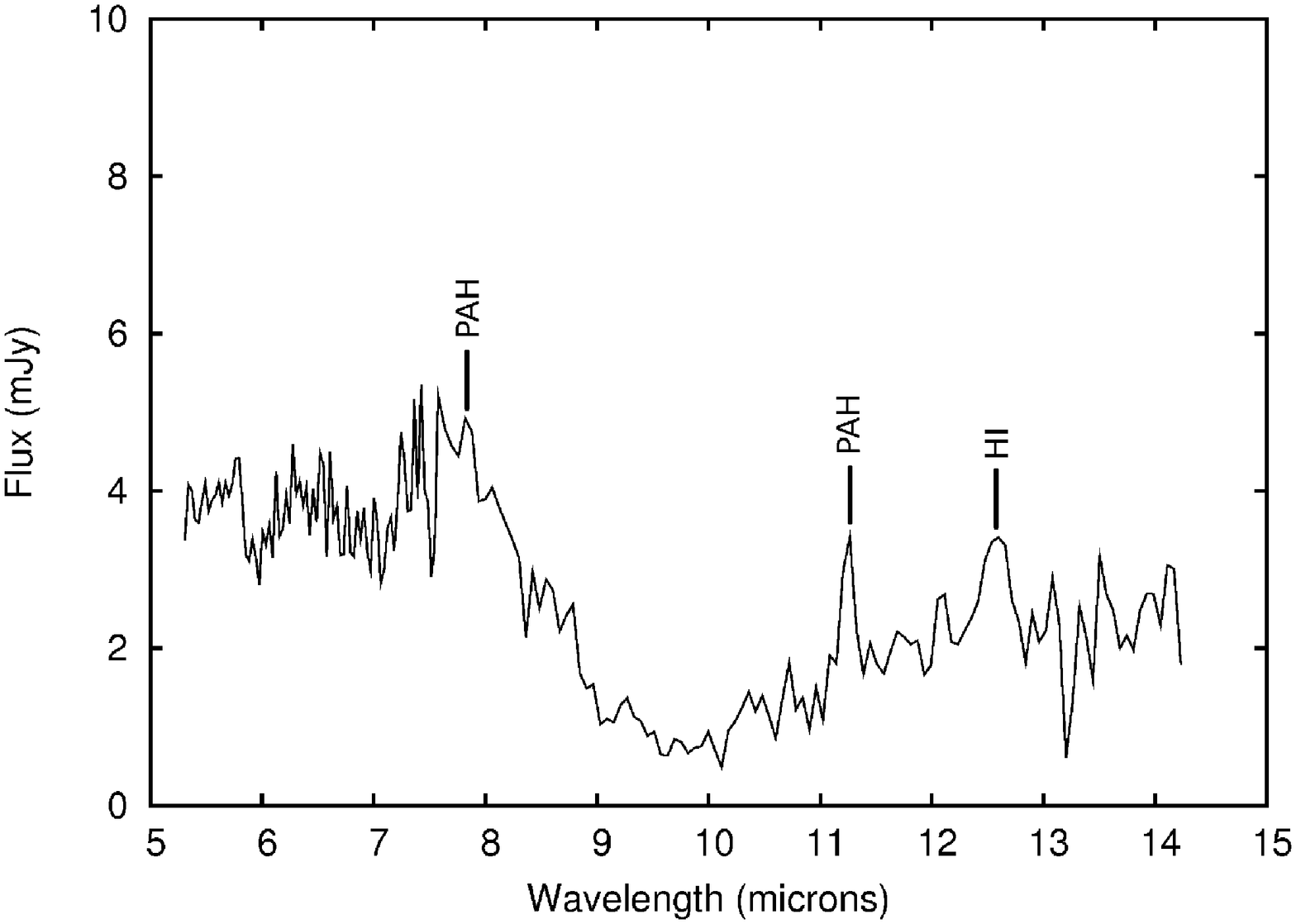}\\
\end{tabular}
\caption{\small IRS spectra of \grs1915~from 5.20 to 14.50~\mic. All detected features are marked. For the MJD~53511 spectrum, the arrows mark the position of the undetected 7.7 and 11.25~\mic\ PAH features.}
\label{spec1915}
\end{center}
\end{figure*}

Spectroscopy was performed with IRS using the SL2 ($5.20-~7.70$~\mic), 
SL1 ($7.40-14.50$~\mic), LL2 ($14.00-21.30$~\mic) and LL1 ($19.50-38.00$~\mic) modules 
with the IRS Peak-up option for a better pointing accuracy. Total exposures times were set to 120~s, divided in 2 sub-exposures 
in SL1 and SL2, and 300~s, divided in ten sub-exposures in LL1 and LL2.  BCD data were reduced following the standard 
procedure given in the IRS Data Handbook
\footnote{http://ssc.spitzer.caltech.edu/irs/dh/dh32.pdf}. The basic steps were bad pixel correction, sky subtraction, as well as 
extraction and calibration (wavelength and flux) of the spectra $-$ with the \textit{Spitzer IRS Custom Extraction } software {\tt Spice} v2.1.2 $-$ for each nod. Spectra were then nod-averaged to improve the signal-to-noise ratio (SNR).  In SL1 and SL2, \grs1915~was always detected with SNRs good enough to allow 
spectroscopic features identification (SNR$\ge$10), as shown in Figure~\ref{spec1915}. Unfortunately, in LL1 and LL2, we never managed to detect the source.

\subsection{\textit{RXTE} observations}

The \rxte\ data were reduced with the {\tt LHEASOFT} v.~6.7.
All data products were extracted from user's good times intervals (GTI).
GTIs corresponded to times when the satellite elevation was greater than 
10$^\circ$ above the Earth limb, the offset pointing less 
than 0.02$^\circ$, and proportional counter unit \#2 was active.
In order to identify the classes, 
we extracted 1s resolution light curves in the 2--60 keV range and 
in the three energy bands defined in \citet{2000Belloni}, 
from the Proportional Counter Array (PCA).  These bands are  2.0$-$5.7~keV 
(channels 0$-$13, PCA epoch 5), 5.7$-$14.8~keV 
(channels 14$-$35), and $>$14.8~keV (channels 36$-$255). The colours were defined as 
HR1=5.7$-$14.8/2.0$-$5.7~keV and HR2=14.8$-$60.0/2.0$-$5.7~keV. The shift of gain between the different 
epochs of PCA leads to different absolute values of the count rates, hardness ratios (HR) and 
position in the colour-colour (CC) diagrams, but the general shape of a given class is easily 
comparable to those of epoch 3 \citep{2000Belloni}, and therefore allowed us to easily identify 
the variability class in each observation \citep[see][]{2008Rodrigueza}. 16~s resolution light curves
were extracted from standard 2 data between 2.0 and 18.0~keV.
All these light curves were corrected for background, using the latest PCA background models 
available for bright sources. Spectra, including those showing clear spectral variations, were then extracted and averaged over the entire observations for a better spectral fitting.

\subsection{\textit{INTEGRAL} observations}

Our study makes use of the \intl\ Soft Gamma-Ray Imager \citep[ISGRI,][]{2003Lebrun}
$-$ the low energy detector of the Imager On-Board \intl\ (IBIS) $-$ to cover the 18.0 
to $\sim300.0$~keV energy range, and the X-ray monitors JEM$-$X \citep{2003Lund} to cover the 3.0$-$30.0~keV one. 
The data reduction process can be found in \citet{2008Rodrigueza, 2008Rodriguezb}.

\subsection{MIR data dereddening}

The MIR data were dereddened from an optical absorption $A_{\rm V}\,=\,19.6\pm1.7$ \citep{2004Chapuis} using the extinction laws given in \citet{2006Chiar}. In their paper, the authors derived the $A_{\rm \lambda}/A_{\rm K}$ ratio rather than the usual $A_{\rm \lambda}/A_{\rm V}$ ones. To express the absorption ratio in a standard way, we assigned to $A_{\rm K}$ the value derived using the extinction law given in \citet{1999Fitzpatrick} for the diffuse ISM ($R_{\rm V}\,=\,3.1$), which is $A_{\rm K}\,=\,0.111\times A_{\rm V}$. Although it has recently been demonstrated that the long-believed 
universal NIR extinction curve showed sharp variations depending on the line-of-sight \citep[][and references therein]{2009Fitzpatrick}, these variations are rather observed outside of the Galactic plane (Galactic centre excluded), and our choice to fix $R_{\rm V}$ to the value of the diffuse ISM has no strong incidences in the dereddening of the \grs1915\ data. 

Moreover, if the authors give a universal expression up to 8.00~\mic, they then propose two laws that take silicate absorption at 9.70 and 18.00~\mic\ into account. Both differ in the sense that one is valid for the diffuse ISM and the other one for the Galactic centre. Our absorbed MIR spectra exhibit a strong silicate absorption feature at 9.70~\mic\, and 
we tried to deredden each of them using both laws. The best results were systematically obtained with the modelling of interstellar extinction due to the diffuse ISM, the silicate feature in the Galactic centre being too strong. 

\section{Results}

\subsection{Light curves}

\begin{figure}
\begin{center}
\includegraphics[height=15cm,angle=270]{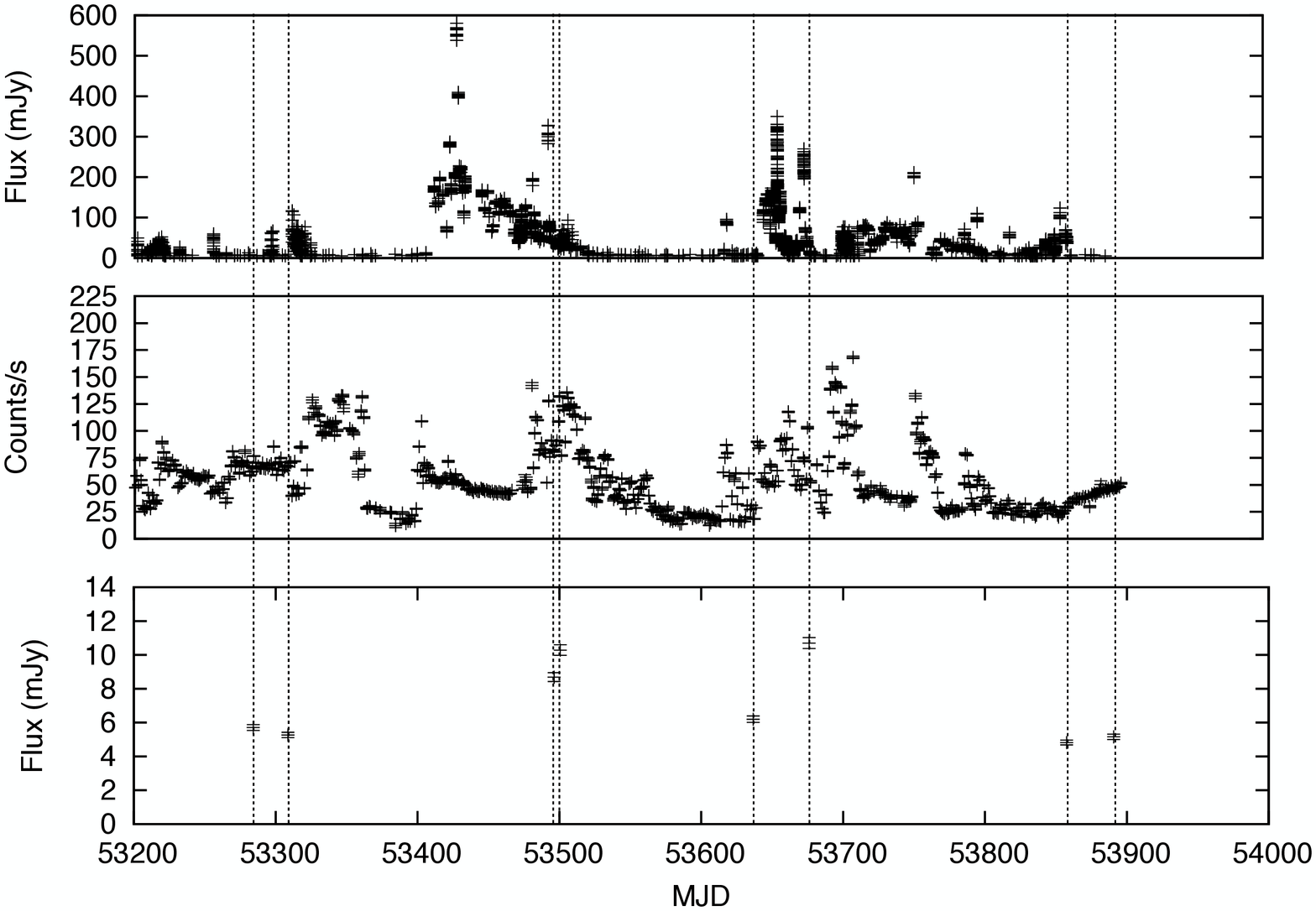}
\caption{\small Ryle (15~GHz, top),  ASM (1.2$-$12.0~kev, middle), and IRAC (3.59~\mic, bottom) light curves of \grs1915\ between 
MJD~53200 and MJD~53900.}
\label{lc}
\end{center}
\end{figure}

The fluxes listed in Table~\ref{fluxirac1915} show that \grs1915\ is strongly variable in all the IRAC filters, which confirms the variation detected by \citet{2003Fuchsa}.  The first step of our study was to compare the evolution of the source's MIR emission with its X-ray and radio ones. Fig.~\ref{lc}\ displays the 3.59~\mic\ (IRAC), 1.2$-$12~keV (ASM) and 15~GHz (Ryle) light curves of \grs1915\, covering the time interval from 
MJD~53200 to MJD~53900. Although we cannot claim for a correlation in the existing data, it is noticeable that the source is at its minimum level of MIR emission (or rather the minimum level of our measurements) when the microquasar has a steady and relatively low X-ray activity (around MJD~53280,  MJD~53308, MJD~53857, MJD~53890). On the other hand, the maximum level of its MIR emission was measured when the source entered in a quite high and unstable X-ray activity, around MJD~53500 and MJD~53676. This behaviour could be a hint for an X-ray/MIR correlation.

Moreover, the comparison with the Ryle telescope light curve could point towards a bimodal relation between the radio and the MIR activities. Indeed, each time \grs1915\ is detected at 15~GHz with a flux higher than 22~mJy (MJD~53496, 53500, and 53676), the MIR emission appears to be at its highest, while it is at its lowest when the radio flux is less than 1~mJy (see Fig.~\ref{lc}\ and Table~\ref{logobs}). Nevertheless, it is important to point out that during all our spectroscopic and photometric observations, \grs1915\ has never been found in the \textit{plateau} state; a MIR emission from the compact jets can consequently be excluded from now on. Each time the Ryle flux is high, the source is rather in the decaying phase of a giant discrete ejection, whose emission is optically thin. This MIR/radio bimodal relation is therefore very likely related to the disc activity.

\subsection{\rxte/\intl\ SEDs}

The quasi-simultaneous high-energy spectra were fitted in {\tt ISIS} v.1.5.0, between 
3.0 (6.0) and 20.0~keV for PCA (JEM$-$X), as well as 20.0 and 200.0~keV for 
ISGRI and HEXTE. In most cases, they were further rebinned so as to obtain good quality spectra. We added 
3 \% systematic errors to the uncertainties in the JEM-X spectra, 2\% to the ISGRI ones and 1\% to the PCA ones 
before the fitting process. In all fits a normalisation constant was added to account 
for uncertainties in the cross calibration of the instruments.

The model we used during the fitting processes consisted in the combination of a multicolour black body \citep[\textit{diskbb},][]{1984Mitsuda}, accounting for the accretion disc emission, a comptonisation component \citep[\textit{comptt},][]{1994Titarchuk} for the corona and a \textit{Gaussian} for the iron feature at 6.4~keV when present, all modified by photo-electric absorption (\textit{phabs}). The column density was fixed at 3.5$\times10^{22}$~atoms~cm$^{-2}$, as measured by \citet{2004Chapuis} from radio observations. If this value is consistent with some measurements derived from X-ray observations \citep[see \textit{e.g.}][]{1998Ebisawa, 2006Mcclintock}, several authors derived higher column densities clustered in the range 5$-$8$\times10^{22}$~atoms~cm$^{-2}$ \citep[see \textit{e.g.}][]{2002Klein, 2002Lee}. Nevertheless, our attempts with higher values always gave equally good fits and did not change significantly the parameters of the disc and the comptonisation components. Our goal being to fit broad band SEDs including both X-ray and MIR data, we decided to use the same column density value in both spectral domains. 

A discussion on the validity of the use of \textit{comptt} to model the hard X-ray emission can be found in \citet{2008Rodriguezb}. Following the authors, the optical depth was fixed to 0.01,  the lowest allowed value, when \grs1915\ was in the states A or B, during which the hard X-ray emission is better fitted by a power law. We made this choice of \textit{comptt} because the use of a power law strongly overestimates the contribution of the corona at low energy and consequently underestimates the disc's one, which has consequences in the optical and infrared domains. Moreover, a power law diverges once the data are dereddened, which forbids a spectral fitting from the X-rays to the MIR. $comptt$ on the contrary peaks at $3kT_{\rm disc}$ and is negligible in the optical and the infrared.

Finally, the Gaussian width was fixed at 0.8~keV and the iron line feature's energy was allowed to vary between 5.0 and 7.0~keV. Table~\ref{par1915he} gives the best-fit parameters found for the spectra obtained on MJD~53280, MJD~53284, MJD~53500, MJD~53676, MJD~53851, and MJD~53890: $kT_{\rm disk}$ and $Norm$ are the accretion disc's temperature and norm, and $kT_{\rm e}$ and $\tau$ are the hard component's electrons temperature and opacity.
The spectra obtained on MJD~63660 and MJD~53661 did not need any disc component but an extra power law. Their best-fit parameters are listed in Table~\ref{par1915he2}.

\subsection{X-ray to MIR SEDs:  the photometric case}

The first step to model the MIR emission of \grs1915\ was to understand in which extent the IRAC fluxes 
could be explained by the Rayleigh-Jeans tail of the accretion disc combined with the stellar emission. We then 
built the X-ray to MIR SEDs of \grs1915\ with the \rxte/\intl\ and IRAC data obtained quasi-simultaneously on 
MJD~53284 and MJD~53890, i.e. when the measured photometric fluxes of the source were at their lowest. Concerning the latter date, it is worth noting that about 1.2~days separate the IRAC data from the \rxte\ ones. Nevertheless, the light curve displayed in Fig.~\ref{lc} shows that between MJD~53890 and MJD~53892, the source had a steadily increasing 1.2$-$12.0~keV X-ray emission (with no flares) and that the disc parameters were therefore barely variable.

\begin{figure}
\begin{center}
\begin{tabular}{c}
MJD~53284\\
\includegraphics[width=7.5cm]{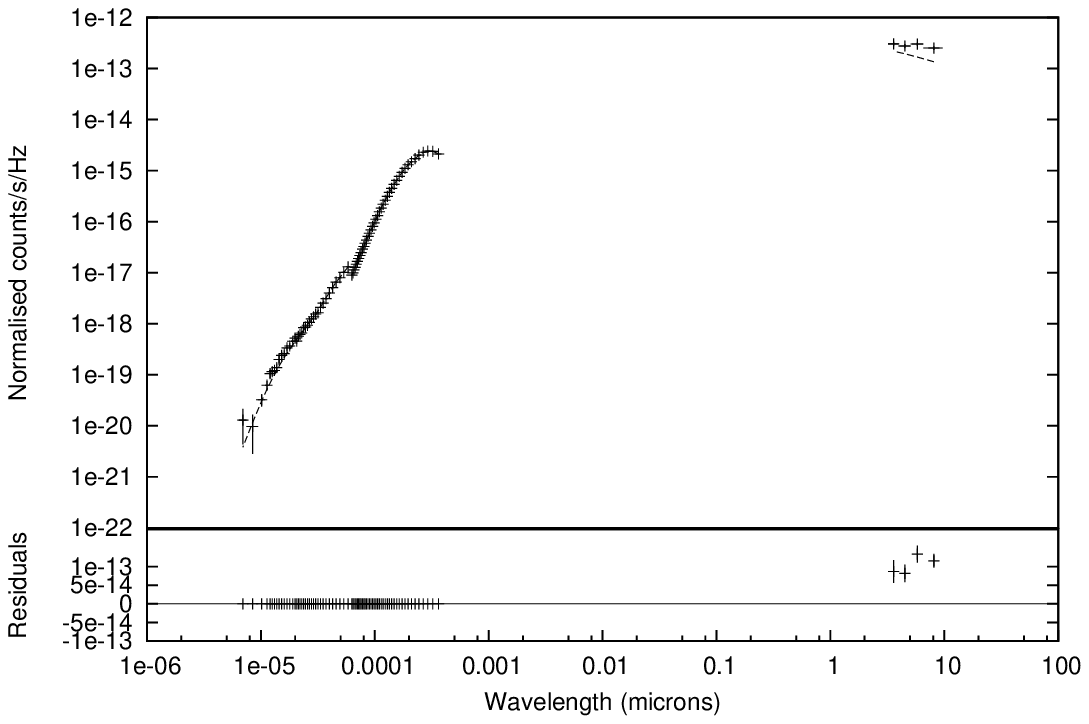}\\
\\
MJD~53890\\
\includegraphics[width=7.5cm]{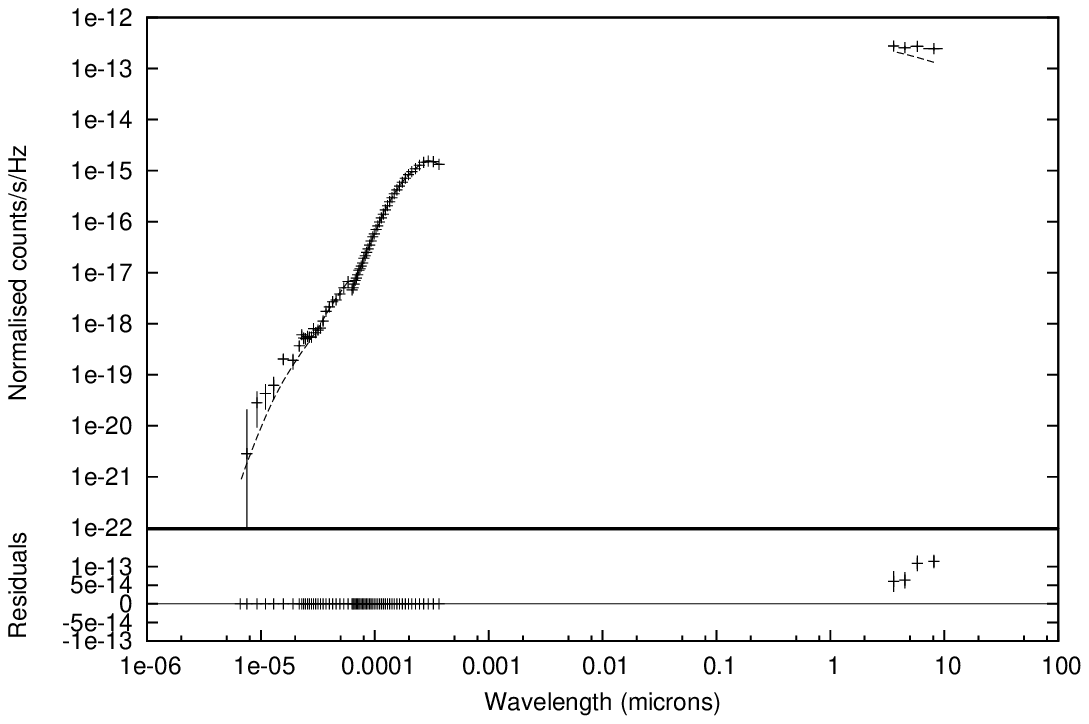}\\
\end{tabular}
\caption{\small \grs1915\ X-ray to MIR SEDs $-$ fitted with the model \textit{diskbb+gaussian+comptt+bbodyrad} $-$ built with the MJD~53284 and MJD~53890 \rxte/PCA+HEXTE and IRAC data. A MIR excess is detected.}
\label{mirexcess}
\end{center}
\end{figure}

For each SED, the dereddened fluxes were stored in an ASCII file. They were then fitted along with the quasi-simultaneous high-energy data  into {\tt ISIS},  combining a spherical black body (\textit{bbodyrad}) and the model described above. \citet{2001Greinera} showed that the companion star of \grs1915\ was a K/M giant whose temperature is about $T_\ast\,=\,4800_{-500}^{+300}$~K,  and \citet{2004Harlaftis} found that its radius was around 19~\rsun. During the fitting process, we then considered two distinct cases: 

\begin{itemize}
\item first, the temperature was allowed to vary between 2800 to 5000~K, which is the temperature scale of K/M giant stars \citep[see \textit{e.g.}][]{1999Vanbelle}. We obtained good fits, but for temperatures that systematically pegged at the minimum allowed, and for stellar radii clustered in the range 38$-$77~\rsun\, depending on the considered distance. This is very unlikely because a giant star with such a low temperature would rather have a radius $R_\ast\,\ge\, 150$~\rsun\ \citep{1998Dumm, 1999Vanbelle},
\item second, the star's temperature and radius were fixed to those of a K2 giant star, i.e. 4520~K and 21~\rsun\ as given in \citet{1999Vanbelle}, and the source's distance was allowed to vary between 6 and 12~kpc. In both cases, the best fits, obtained for a distance of 6~kpc, were unable to completely reproduce the IRAC fluxes as there always was a MIR excess (see  the fits displayed in Fig.~\ref{mirexcess}). 
\end{itemize}

We then conclude that even when it is at its lowest, and in absence of any radio activity, the MIR emission of \grs1915\ cannot be explained only by the companion star and Rayleigh-Jeans tail of the accretion disc. Moreover, on MJD~53500 and MJD~53676, the IRAC fluxes of the source almost doubled compared to their lowest values. It is impossible that this increase comes from the companion star. 
On the contrary, the disc dominates the 3.0$-$200.0~keV unabsorbed X-ray emission, which was not the case on MJD~53284 and MJD~53890 (Table~\ref{par1915he}). This therefore suggests a relation between the MIR and the disc activities of \grs1915. 

\section{Origin of the MIR excess of \grs1915\ }

Although we excluded contributions from the companion star and the compact jets, the MIR excess we detected as well as the increase of the MIR fluxes of \grs1915\ may still be explained by (1)~the presence of a dust component, maybe heated by the companion star and/or the X-ray/UV emission, 
(2)~thermal \textit{bremsstrahlung} from the accretion disc's winds,
(3)~illumination of the accretion disc,
(4)~optically thin synchrotron from a discrete ejection.

\subsection{A photoionised dust component?}

All the IRS SL1/2 absorbed spectra of \grs1915\ are displayed in Fig.~\ref{spec1915}, and Table~\ref{raie1915} lists all the features 
we detected. All the measurements were carried out using the task {\tt IDEA} of the data reduction software {\tt SMART}~v.~6.4.0.  
The central wavelength $\lambda_{\rm fit}$, equivalent width $\mathring{W}$, full-width at half-maximum \textit{FWHM}, flux, and signal-to-noise ratio \textit{SNR} of each feature were computed through gaussian fitting. The highest source of error on the equivalent width and line flux measurement is due to the continuum, particularly uncertain at these wavelengths. Each time, instead of trying to fit it globally, the continuum was assessed in the vicinity of each feature through a linear function fitting. Several attempts showed that the resulting systematic errors were about 5\% of the measured flux, which were quadratically added to the statistical uncertainties.

Along with several \ion{H}{1} and \ion{H}{2} emission lines that likely originate from the accretion disc, and a strong silicate absorption feature, we detected in each spectrum but the one obtained on MJD~53511, the so-called unidentified infrared features at 7.70~\mic\ (with 2 primary components at 7.60 and 7.80~\mic) and 11.25~\mic\ (detected between 11.20 and 11.30~\mic). These lines are thought to be created by the family of the polycyclic aromatic hydrocarbon molecules \citep[PAH,][]{1984Leger, 1985Puget, 1985Allamandola}, which are found in the MIR spectra of many objects with associated dust and gas components illuminated by UV photons. 

The presence of such features in the MIR spectrum of \grs1915\ infers that there is dust in the system. Moreover, despite the strong uncertainties, all the PAH lines appear to be strongly variable in flux; this might be related to the photoionisation state of the environment of \grs1915. Indeed, the excitation mode of PAH molecules is vibrational; the 7.70~\mic\ feature is due to C$-$C stretching, whereas the 11.25~\mic\ is due to C$-$H out-of-plane bending. The C$-$H bond being weaker than the C$-$C one, the photoionised PAH is principally detected at 7.70~\mic. The flux ratio  $F_{7.7}/F_{11.3}$ is therefore a good indicator of the photoionisation state of the environment in which PAH molecules are detected, and the larger it is, the higher the photoionisation degree \citep[see \textit{e.g.}][for comprehensive reviews on PAH properties]{1989Allamandola, 2008Tielens}.

\begin{figure}
\begin{center}
\includegraphics[width=6.5cm, angle=270]{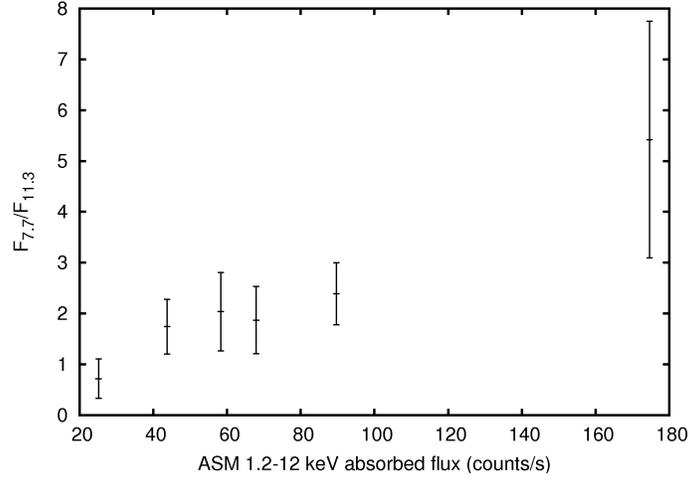}
\caption{\small Ratio of the PAH lines absorbed fluxes at 7.70 and 11.25~\mic~in function of the 1.2-12.0~keV ASM absorbed flux. Errorbars are given at 3$\sigma$.}
\label{xpah}
\end{center}
\end{figure}

The companion star of \grs1915\ is a late K giant. If its emission can heat up a potential dust component, it is impossible that it photoionises it as it is too cold to emit enough energetic UV photons. Therefore, only the X-ray emission can be responsible for the photoionisation of the PAH molecules. To validate this hypothesis, we compared, for each spectrum, the evolution of the $F_{7.7}/F_{11.3}$ PAH flux ratio with the corresponding 1.2$-$12.0~keV X-ray flux of \grs1915. The result is displayed in Fig.~\ref{xpah}. Despite the strong uncertainties due to the silicate absorption, it suggests a correlation; the dust may therefore be photoionised by the X-ray/UV photons originated from the accretion disc and/or the corona. This would partly explain why no PAH molecules are detected in the \grs1915\ MIR spectrum obtained on MJD~53511. Indeed, a  [\ion{Ne}{2}] emission feature is present at 12.81~\mic. The Ne atom ionisation potential is about 21.56~eV, while the PAH one is smaller than 10~eV, depending on the molecule size \citep{2005Ruiterkamp}. The absence of the 11.25~\mic\ PAH feature could consequently mean that almost all the PAH was photoionised. But this spectrum also displays a huge increase of the continuum below about 9.00~\mic, reaching the flux level measured through photometry on MJD~53500 i.e. about 8.00~mJy and 5.00~mJy at 5.80~\mic\ and 8.00~\mic, respectively. The reason for this increase will be discussed in the next section, but the non-detection of the PAH feature at 7.70~\mic\ could be due to a contamination by the continuum, as the peak flux of the feature appears to be less than 6.00~mJy in all the other spectra where it is detected.

\subsection{Effect of the thermal \textit{bremsstrahlung}}

Thermal \textit{bremsstrahlung} from an expanding wind might be partially responsible for the MIR emission of \grs1915.  Indeed, 
\citet{1983Begelman} showed that X-ray driven winds could form above an accretion 
disc heated by X-ray radiation with luminosity a few percents above the Eddington limit, which is likely the case of  
\grs1915\ as \citet{2002Lee} and more recently \citet{2009Neilsen} and \citet{2009Ueda} detected such wind in the system. 

Thermal \textit{bremsstrahlung} was already invoked in \citet{1994Paradijsa} to explain the strong MIR excess the authors had detected in the emission of GRO~J0422+32. In their paper, they used the formalism given in \citet{1979Rybicki} to assess the expected 10.80~\mic\ luminosity of a spherical expanding wind. We followed the same steps, for a disc wind emitted in a solid angle $\Omega$, to assess the expected monochromatic luminosity of \grs1915\ due to free-free emission at 8.00~\mic. In such a wind, the mass-loss rate $\dot{M_{\rm w}}$ is:
\begin{equation}
\dot{M_{\rm w}}\,=\,\left ( \frac{\Omega}{4\pi} \right ) \times 4\pi r^2 m_{\rm p } n_{\rm e} v_{\rm w}
\label{mp}
\end{equation} 

where $r$ is the distance within the wind, $m_{\rm p }$ the proton mass, $n_{\rm e}$ the electronic density, and $v_{\rm w}$ the wind velocity. Following \citet{1979Rybicki}, the thermal \textit{bremsstrahlung} emissivity at the frequency $\nu$ can be written as:
\begin{equation}
\epsilon_\nu \,=\, 6.80 \times 10^{-{45}} \left ( \frac{n_{\rm e}^2} {\sqrt{T}} \right ) e^{-{\frac{h\nu}{k_{\rm B}T}}} \times g\,\,\,\,\,\,\textrm{W~cm$^{-3}$~Hz$^{-1}$}
\label{eps}
\end{equation}
where $T$ is the wind temperature, \textit{g} the Gaunt factor, and $h$ and $k_{\rm B}$ the Planck and Boltzmann constants, respectively. In the following, we fix $g$ to 1, corresponding to a large angle regime in the interaction between an electron and an ion.

Replacing $n_{\rm e}$ from Eq.~\ref{mp} into Eq.~\ref{eps}, and integrating $\epsilon_\nu$ over the radial distance $r$ between the launching radius $R_0$ (in cm) and infinity, we obtained the monochromatic luminosity $L_\nu$ at the frequency $\nu$:
\begin{eqnarray}
L_\nu&=&\left (\frac{\Omega}{4\pi} \right ) \int_{R_0}^{\infty} \epsilon_\nu \times 4\pi r^2 dr\\
&=& 2.04 \times 10^{8} \, \frac{e^{-{\frac{h\nu}{k_{\rm B}T}}}}{\sqrt{T}}  \frac{\dot{M_{\rm w}}^2}{\left (\frac{\Omega}{4\pi}\right)v_{\rm w}^2 R_0}\,\,\,\,\,\,\textrm{W~Hz$^{-1}$}
\end{eqnarray}

Following \citet{2009Ueda}, we can approximate the mass-loss rate $\dot{M_{\rm w}}$ as:
\begin{equation}
\dot{M_{\rm w}}\,\approx\,1.00\times 10^{11}\left(\frac{\Omega}{4\pi}\right)v_{\rm w}\,\,\textrm{kg~s$^{-1}$}
\end{equation}
which in turn gives the following expression for the monochromatic luminosity:
\begin{equation}
L_\nu\,=\,2.04 \times 10^{30} \left(\frac{\Omega}{4\pi}\right)\frac{e^{-{\frac{h\nu}{k_{\rm B}T}}}}{R_0\sqrt{T}}
\end{equation}

In their thermally-driven wind model, \citet{1983Begelman} introduced four important parameters, the 
Compton temperature $T_{\rm C}$, where heating from Compton scattering and cooling from inverse 
Compton are balanced out, the Compton radius $R_{\rm C}$, for which the escape velocity equals 
the isothermal sound speed at the Compton temperature, the critical luminosity $L_{\rm cr}$, above which a Compton 
heating disc wind can overcome gravity, and $T_{\rm ch}$, which is the characteristic temperature of a parcel of gas that rises 
at a height $R_0$ above the disc in a finite heating time. 
Those parameters are defined as:
\begin{eqnarray}
T_{\rm C}&=&\frac{1}{4k_{\rm B}}\frac{\int_{\nu_1}^{\nu_{\rm N}} h\nu L_{\nu} d\nu}{\int_{\nu_1}^{\nu_{\rm N}} L_{\nu} d\nu}\,\textrm{K}\\
R_{\rm C}&=&\frac{9.80\times 10^{17}}{T_{\rm C}}\frac{M_{\rm X}}{M_\odot}\,\textrm{cm}\\
L_{\rm cr}&\approx&2.88\times 10^2 \, \frac{L_{\rm E}}{\sqrt{T_{\rm C}}} \\
T_{\rm ch}&=&T_{\rm C}\left (\frac{L}{L_{\rm cr}}\right )^{\frac{2}{3}}\left(\frac{R_0}{R_{\rm C}} \right)^{-{\frac{2}{3}}}
\end{eqnarray}
where $M_{\rm X}$ is the black hole's mass, $L$ the X-ray bolometric luminosity, and $L_{\rm E}$ the Eddington luminosity. 
On MJD~53284, using the continuum parameters given in Table~\ref{par1915he}, and integrating 
the system's monochromatic luminosity between 1.0 and 1000.0~keV, we find $T_{\rm C}\,\approx\,5.80\times10^{6}$~K, which leads to 
$R_{\rm C}\,\approx\,2.37\times10^{12}$~cm, $L_{\rm cr}\,\approx\,0.12\times L_{\rm E}$, and $L\,\approx\,0.37\times L_{\rm E}$.

\citet{1983Begelman} and \citet{1996Woods} showed that a disc wind could develop for $R_0\,\ge\, 0.2\times R_{C}$ and would get strong for $R_0\,\ge\, R_{C}$.  
It is reasonable to assert that the thermal \textit{bremsstrahlung}  arises from the gravity-free part of the wind, in which $T\,\approx\,T_{\rm ch}\,\le\,T_{C}$. Then, considering $L\,\approx\,3\times L_{\rm cr}$, Eq.~10 leads to $R_0\,\ge\,3\times R_{\rm C}$. Replacing $T$ by $T_{\rm C}$ and $R_0$ by $3\times R_{\rm C}$ in Eq.~6, the 
inferred luminosity at 8.00~\mic, $L_{\rm 8}$ is:
\begin{equation}
L_{\rm {8}}\,\approx\,1.19\times 10^{14}\left(\frac{\Omega}{4\pi}\right)\,\textrm{W~Hz}^{-1}
\end{equation}
On MJD~53284, the unabsorbed luminosity of \grs1915\ at 8.00~\mic\ is about $8.27\pm1.30\times 10^{13}$~W~Hz$^{-1}$. With a covering 
factor of about 5\%, as measured for \grs1915\ in \citet{2009Neilsen}, we deduce a \textit{bremsstrahlung}-induced luminosity at 8.00~\mic\ of about $(5.60\pm0.55)\times10^{12}$~W~Hz$^{-1}$, which is an order of magnitude lower that what we measured. Even for a larger value of the covering factor as high as 0.2 \citep{2000Proga}, we derive about $(2.38\pm0.23)\times10^{13}$~W~Hz$^{-1}$, which is still too low. It is therefore very likely that thermal \textit{bremsstrahlung} from the accretion disc wind barely contributes to the MIR flux of \grs1915. Moreover, Fig.~\ref{spec1915} shows that on MJD~53511, the MIR continuum of the source below about 9~\mic\ increased while it remained almost the same beyond. Yet, 
the spectral signature of free-free emission is a power law with an index from 0 (optically thin) to 2 \citep[optically thick,][]{1975Wright}, 
the wind being in the optically thin regime in the infrared. An increase of the MIR flux due to \textit{bremsstrahlung} would be detected at all wavelengths, which then excludes it as a reason for the MIR brightening.

\subsection{X-ray to MIR SEDs: irradiation of the disc}

In the soft state, reprocessing $-$ in the outer disc $-$ of X-ray and UV photons originated from the inner part likely dominates the UV and optical emission of microquasars \citep{1990Vrtilek, 1992Fukue, 1993Sambuichi, 1994Paradijsb, 1998Hynes, 2000Esin, 2002Hynes}. In the hard state, there might also be a contribution of the reprocessed hard X-ray photons from the corona. Indeed, \citet{2002Ueda} showed that it could even represent about 20$-$30~\% of the \textit{K}-band emission of \grs1915\ in the \textit{plateau} state.  

We showed that the sharp increase of the MIR emission on MJD~53511 $-$ only detected below about 9.00~\mic $-$ was not due to thermal \textit{bremsstrahlung}. Dust heating is also little plausible as the MIR increase would have been detected at all wavelengths. On the contrary, it seems rather consistent with irradiation of the outer disc and/or optically thin synchrotron from a discrete ejection. To confirm this hypothesis, we built $-$ for MJD~53284, MJD~53511, and MJD~53851 $-$ the X-ray to radio SEDs of \grs1915, including the dereddened MIR spectra. Moreover, on MJD~53284 and MJD~53851, \grs1915\ was not detected in the radio domain so we did not include any radio flux in these SEDs. 

\begin{figure}
\begin{center}
\includegraphics[width=6.5cm, angle=270]{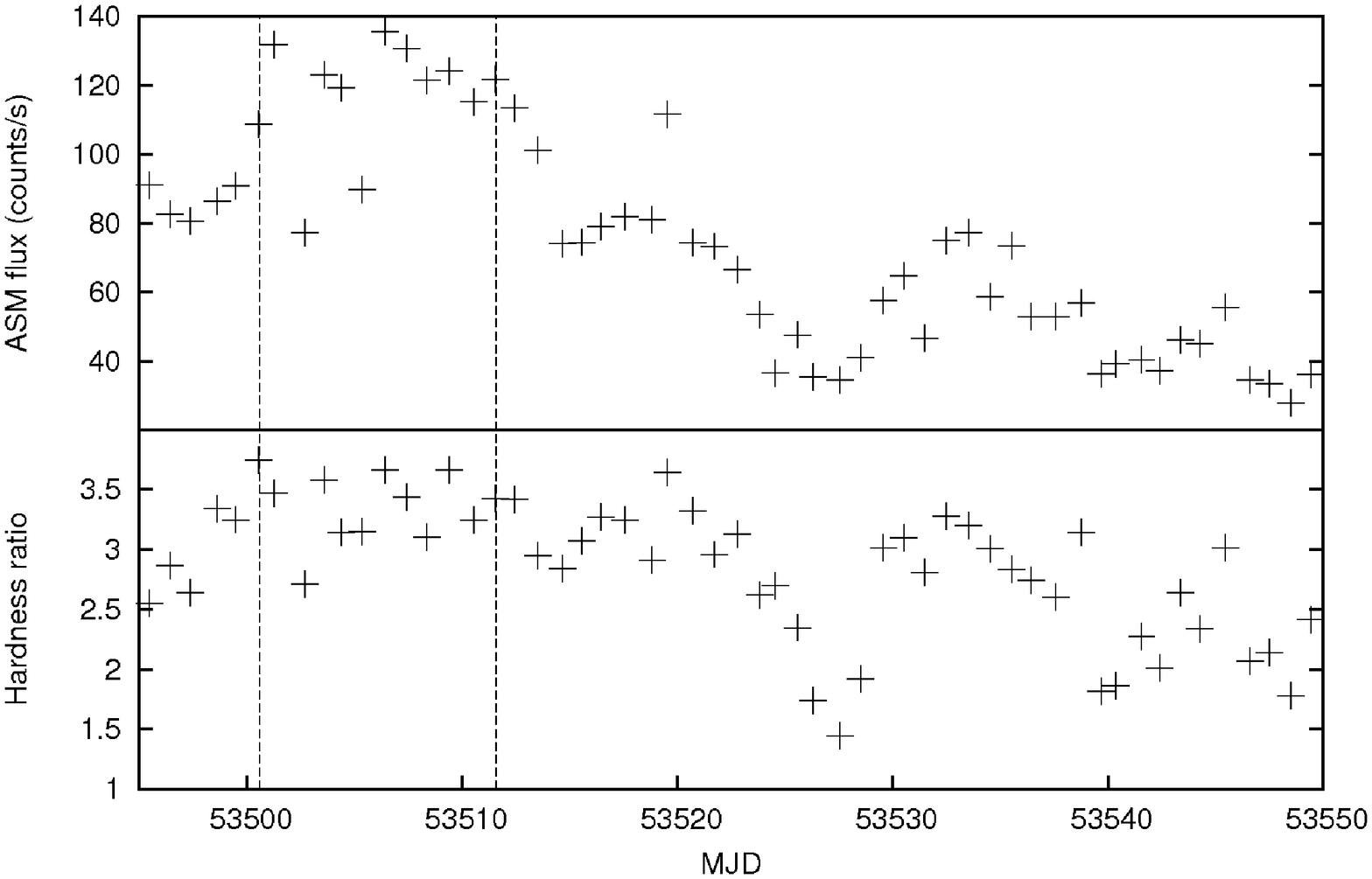}
\caption{\small \textbf{Top}: day-averaged ASM light curve of \grs1915\ between MJD~53495 and MJD~53520.
\newline
\textbf{Bottom}: C/A ASM hardness ratio.}
\label{asm_lite}
\end{center}
\end{figure}

The way we built the MJD~53511 radio to X-ray SED of \grs1915\ deserves some justifications. In the high energy domain, we used the MJD~53500 \rxte\ data because we did not have any high-energy observations quasi-simultaneous with the IRS ones (see Table~\ref{logobs}). Indeed, the ASM fluxes and C/A hardness ratios being similar at both dates (see Fig.~\ref{asm_lite}), it is likely that the disc parameters were the same. Moreover, to compensate the lack of data in the radio domain, we made use of the archival VLA fluxes of the source obtained on MJD~53513 at 8.46, 14.94, and 22.46~GHz (37.90, 23.20, and 16.20~mJy respectively, about 10\% uncertainties)\footnote{\tiny http://www.aoc.nrao.edu/$\sim$mrupen/XRT/GRS1915+105/grs1915+105.shtml\normalsize}. At this epoch, \grs1915\ was in the decaying phase of a giant ejection, and the VLA flux at 14.94~GHz is similar to the ones from the Ryle telescope on MJD~53510 and MJD~53512, proving that the unknown MJD~53511 radio fluxes are barely different from the VLA ones.
\newline

The model we chose to take the reprocessing into account is \textit{diskir} \citep{2008Gierlinski,2009Gierlinski}. Roughly, it is a extension of \textit{diskbb} that includes disc irradiation (both from the inner region and the corona) as well as comptonisation (based on \textit{nthcomp}). The model has nine parameters: the disc's temperature $kT_{\rm disc}$ and norm \textit{Norm} (same as \textit{diskbb}), the hard X-rays power law $\Gamma$ and temperature $kT_{\rm e}$, the ratio between the corona's and the disc's luminosity $L_{\rm c}/L_{\rm d}$, the fraction of hard X-ray emission that illuminates the disc $f_{\rm in}$, the irradiated radius $R_{\rm irr}$ expressed in term of the disc inner radius, the fraction of soft X-ray emission which is thermalised in the outer disc $f_{\rm out}$, and the logarithm of the outer radius $lrout$ expressed in function of the inner radius. The first seven parameters are completely defined by the high-energy data while the two latter are characterised by the optical and infrared ones. 

\textit{diskir} was additively combined to a gaussian accounting for the iron feature at 6.4~keV (0.8~keV frozen width), and we fitted all the data, in {\tt ISIS}, in three steps:
\begin{itemize}
\item we first modified the model with a photo-electric absorption \textit{phabs} ($N_{\rm H}\,=\,3.5\times10^{22}$~atoms~cm$^{-2}$) and fitted the high-energy data only. The electron temperature was frozen to the value found in the previous fits with \textit{comptt} (see Table~\ref{par1915he}), the irradiated radius was frozen to $1.1\times R_{\rm in}$ after several unsuccessful attempts that showed that it was poorly constrained, $f_{\rm in}$ was fixed to 0.1 for the corona-dominated spectrum and to 0.3 for the disc-dominated one \citep[for a fixed 0.1 disc's albedo,][]{1997Poutanen, 2005Ibragimov, 2009Gilfanov}, and $f_{\rm out}$ and $lrout$ were frozen to 0 and 3, respectively, as the X-ray data do not allow to constrain them,
\item we built the new radio to X-ray SEDs with the unabsorbed MIR spectra $-$ stored in ASCII files $-$ the \rxte\ data sets and the VLA radio fluxes (MJD~53511 only), 
\item we finally fitted the global SEDs with the previous model combined to spherical black body component accounting for the companion star emission, another one accounting for the detected dust emission, and a power law for the radio emission (MJD~53511 only). All the parameters were allowed to vary freely, except the stellar ones (temperature and radius fixed to 4520~K and 21~\rsun, respectively),  $kT_{\rm disc}$ and $R_{\rm irr}$ (same values as in step 1), and \textit{lrout} which was frozen to $0.52a$ \citep{2003Chaty}, $a$ being the orbital separation derived from the third Kepler's law for a 30.8~days orbital period \citep{2007Neil}, and a 14~\msun\ and 0.86~\msun\  black hole and companion star, respectively  \citep{2004Harlaftis}. Note that \textit{lrout} was dynamically tied to $R_{\rm in}$ for a 10~kpc distance and a 66$^{\circ}$ inclination.
\end{itemize}

\begin{figure*}
\begin{center}
\begin{tabular}{ccc}
MJD~53511&MJD~53280&MJD~53851\\
\includegraphics[width=3.5cm, angle=270]{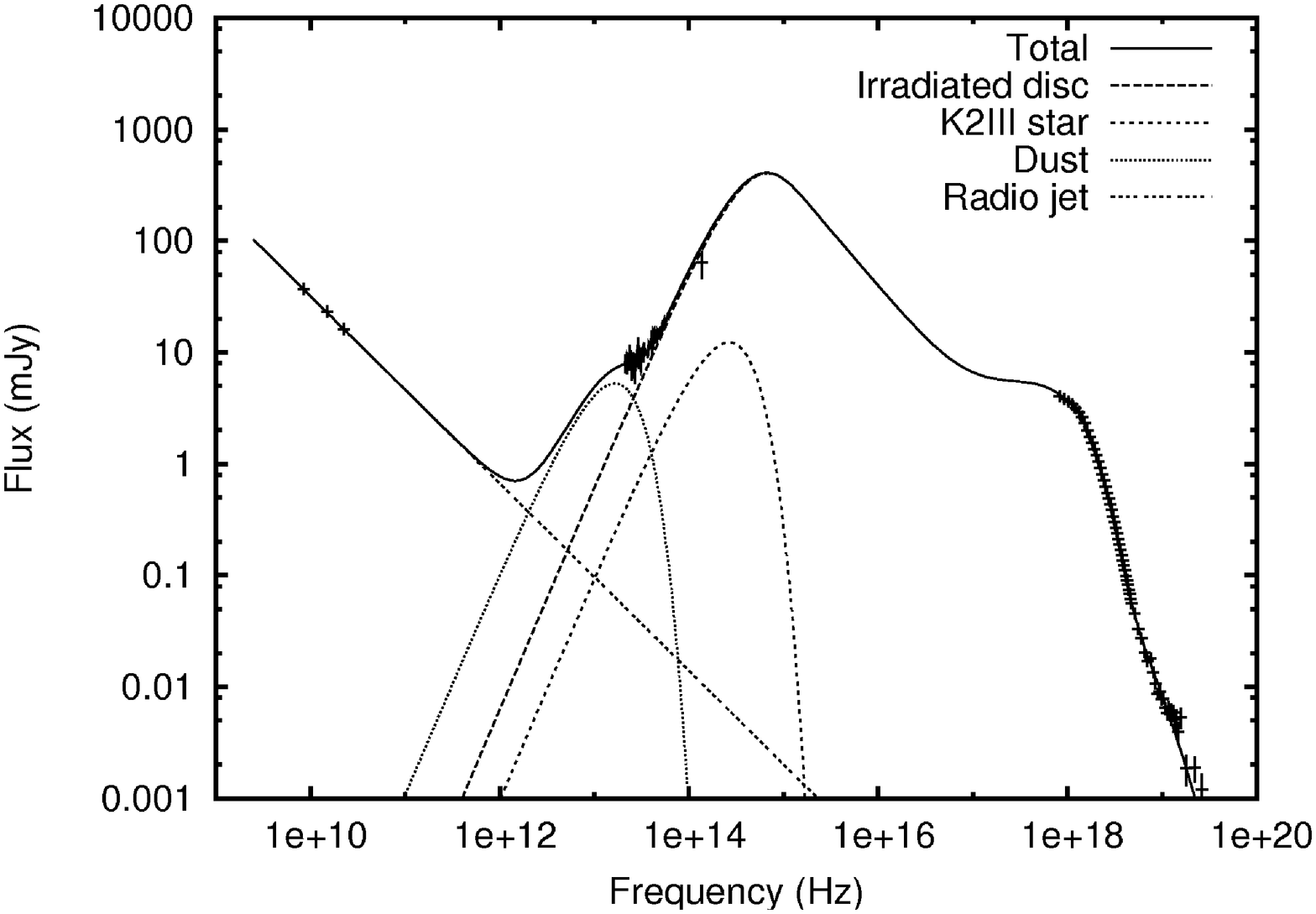}&\includegraphics[width=3.5cm, angle=270]{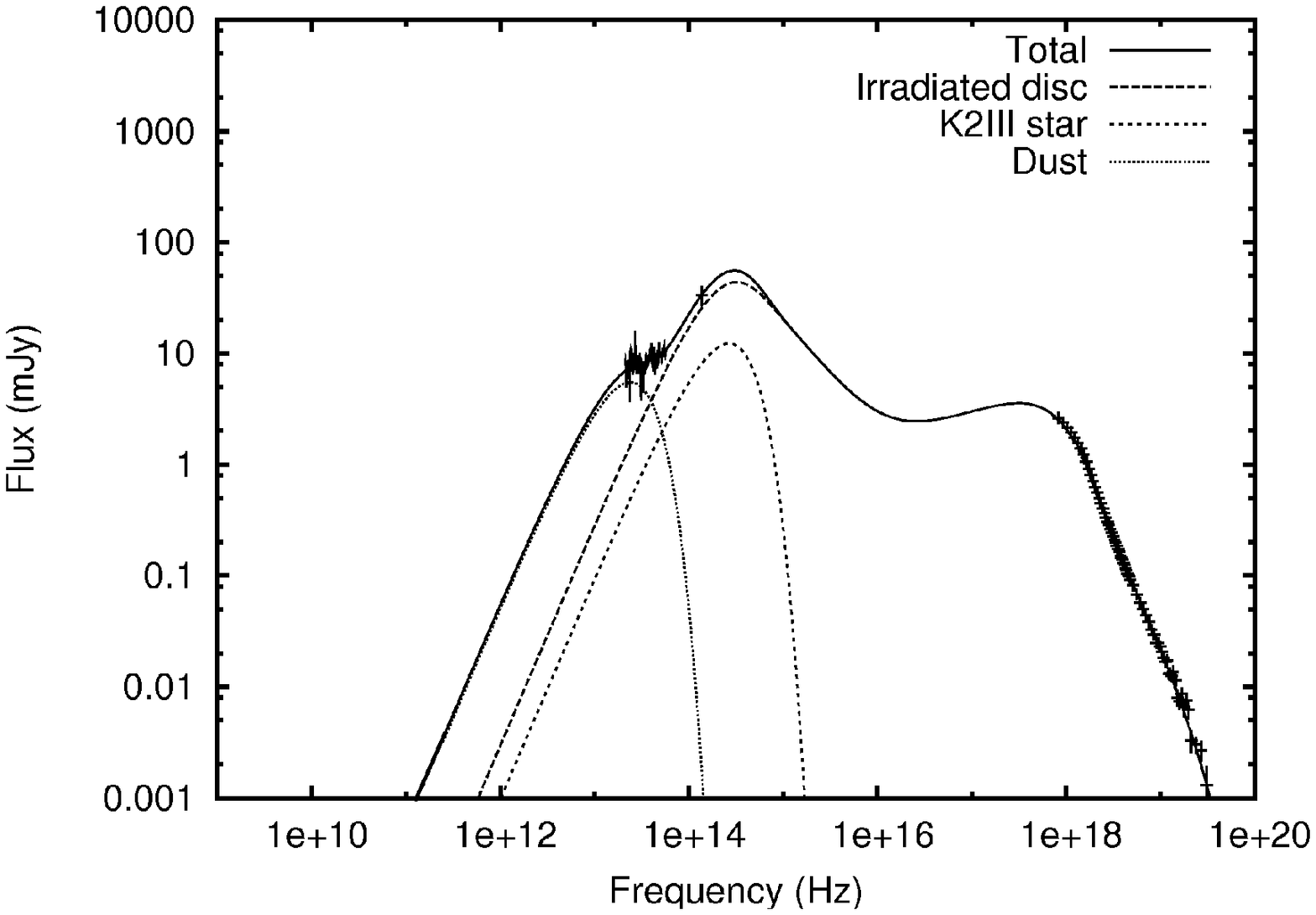}&\includegraphics[,width=3.5cm,
angle=270]{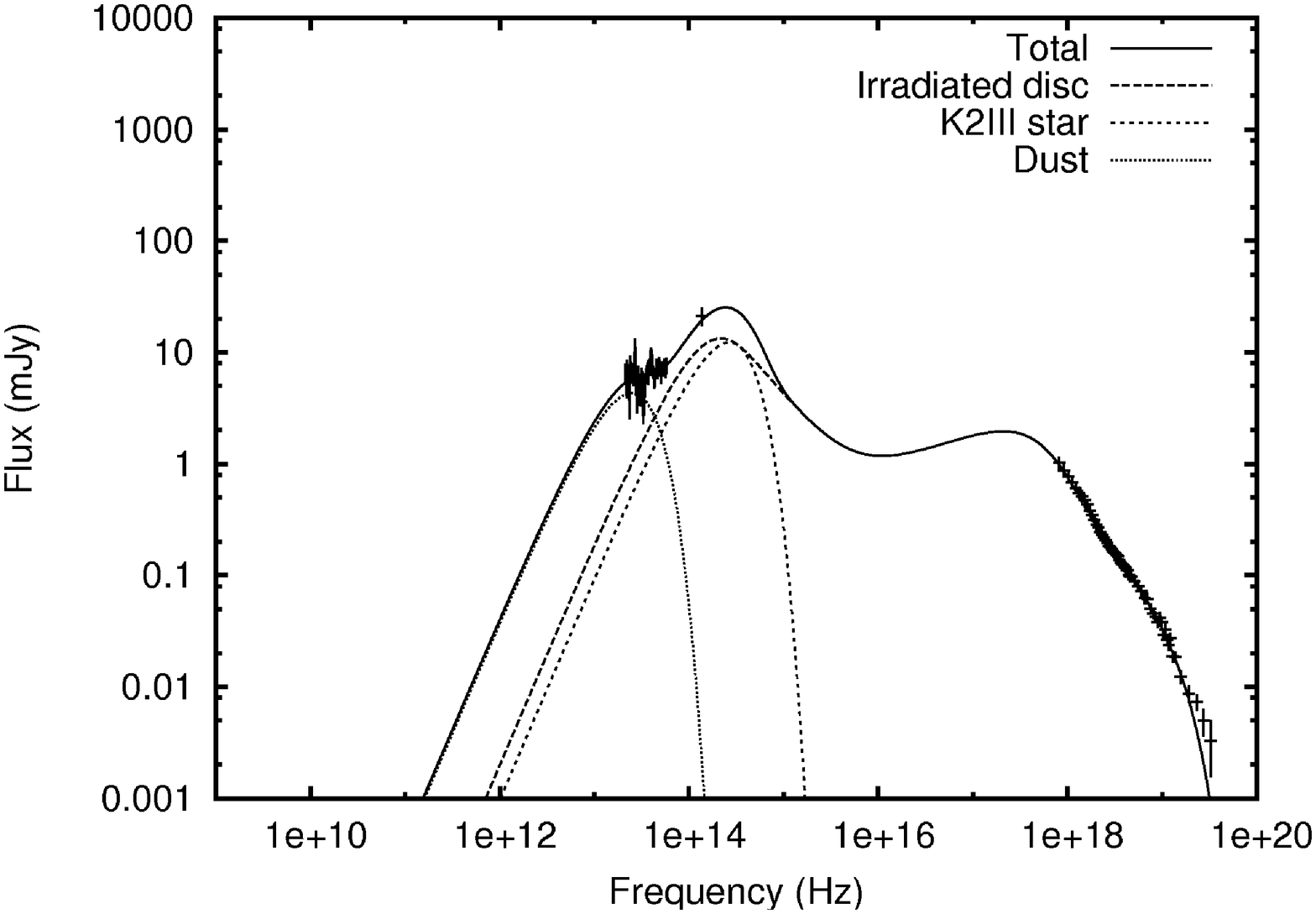}\\
\includegraphics[width=3.5cm, angle=270]{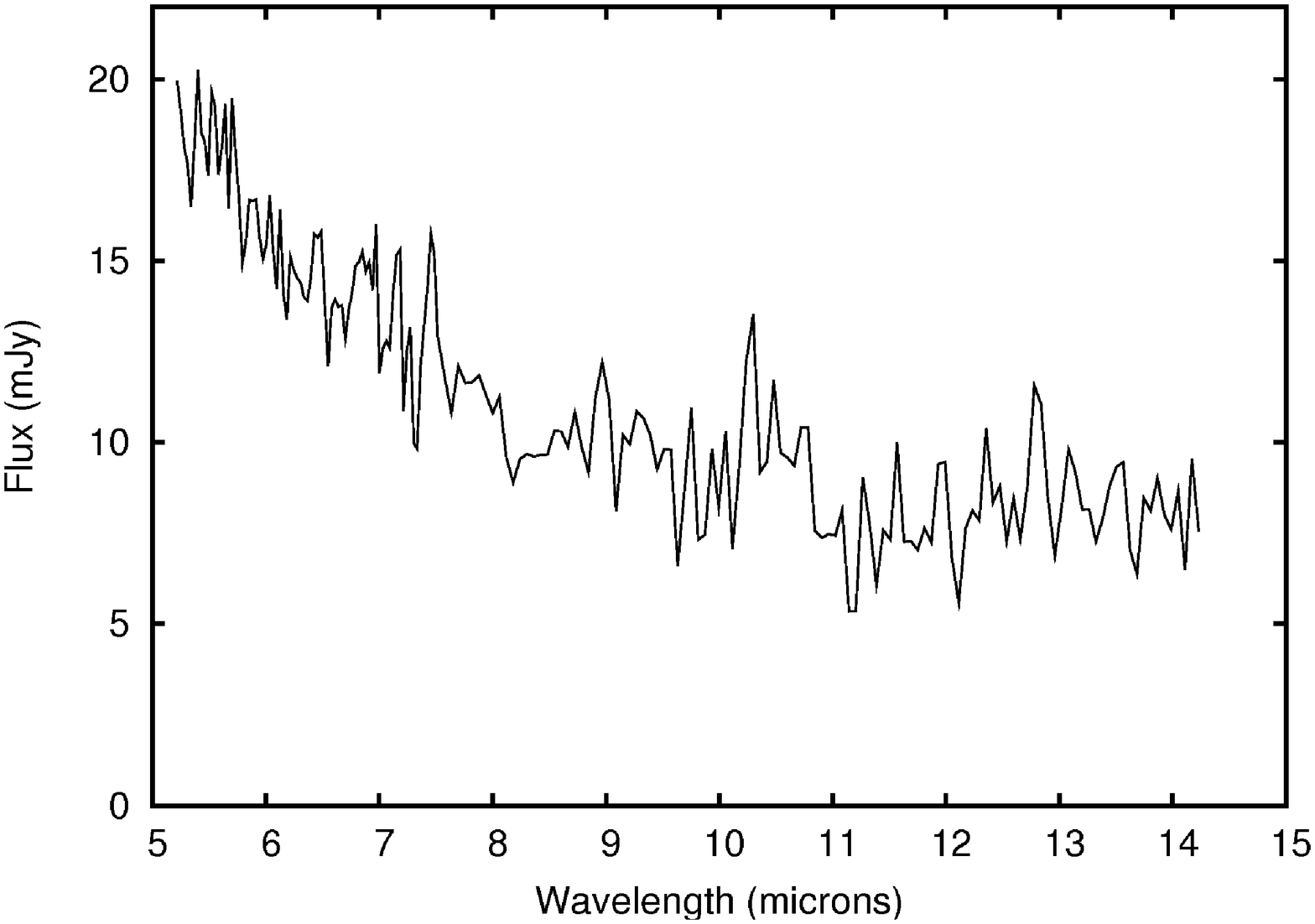}&\includegraphics[width=3.5cm, angle=270]{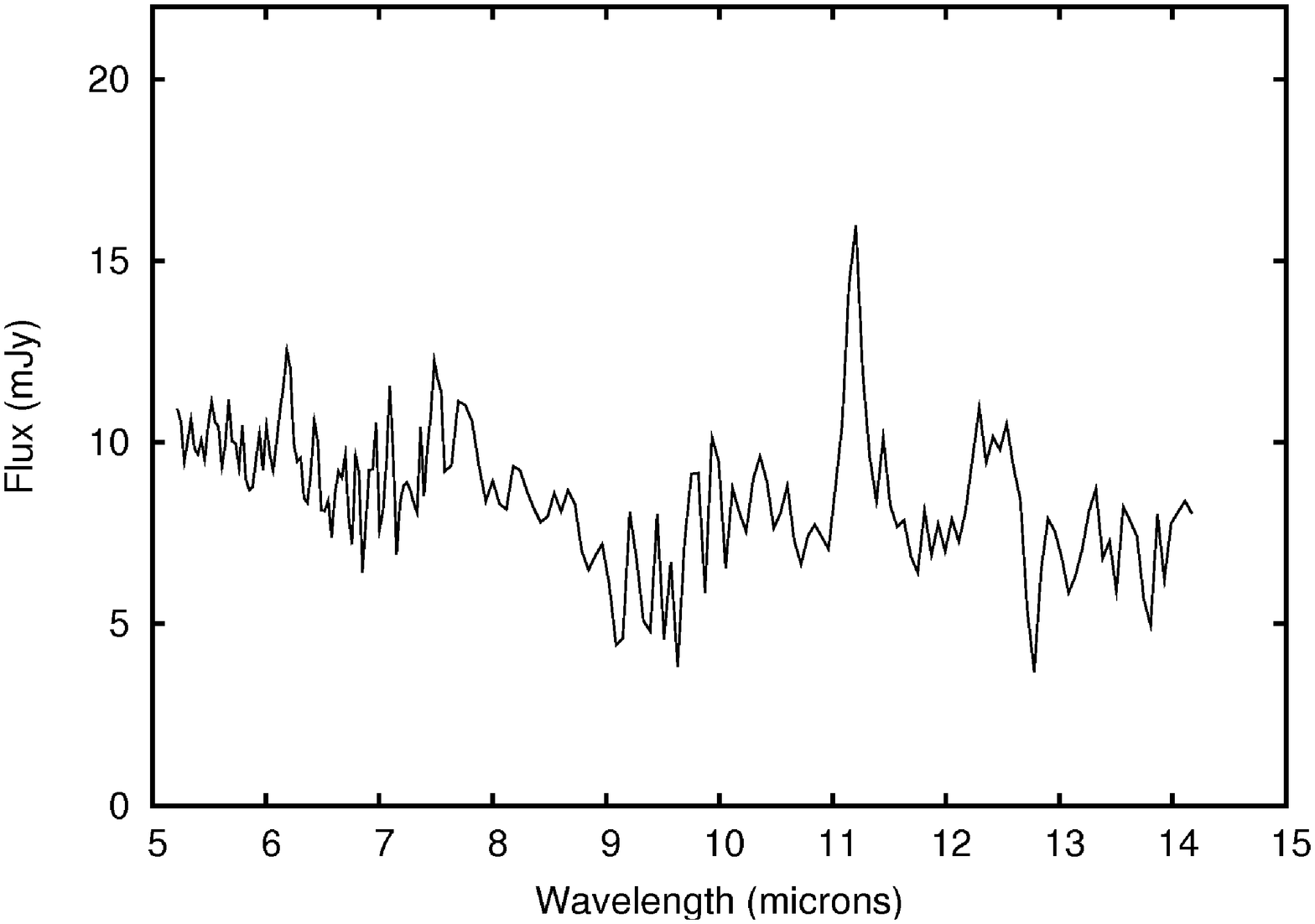}&\includegraphics[,width=3.5cm, angle=270]{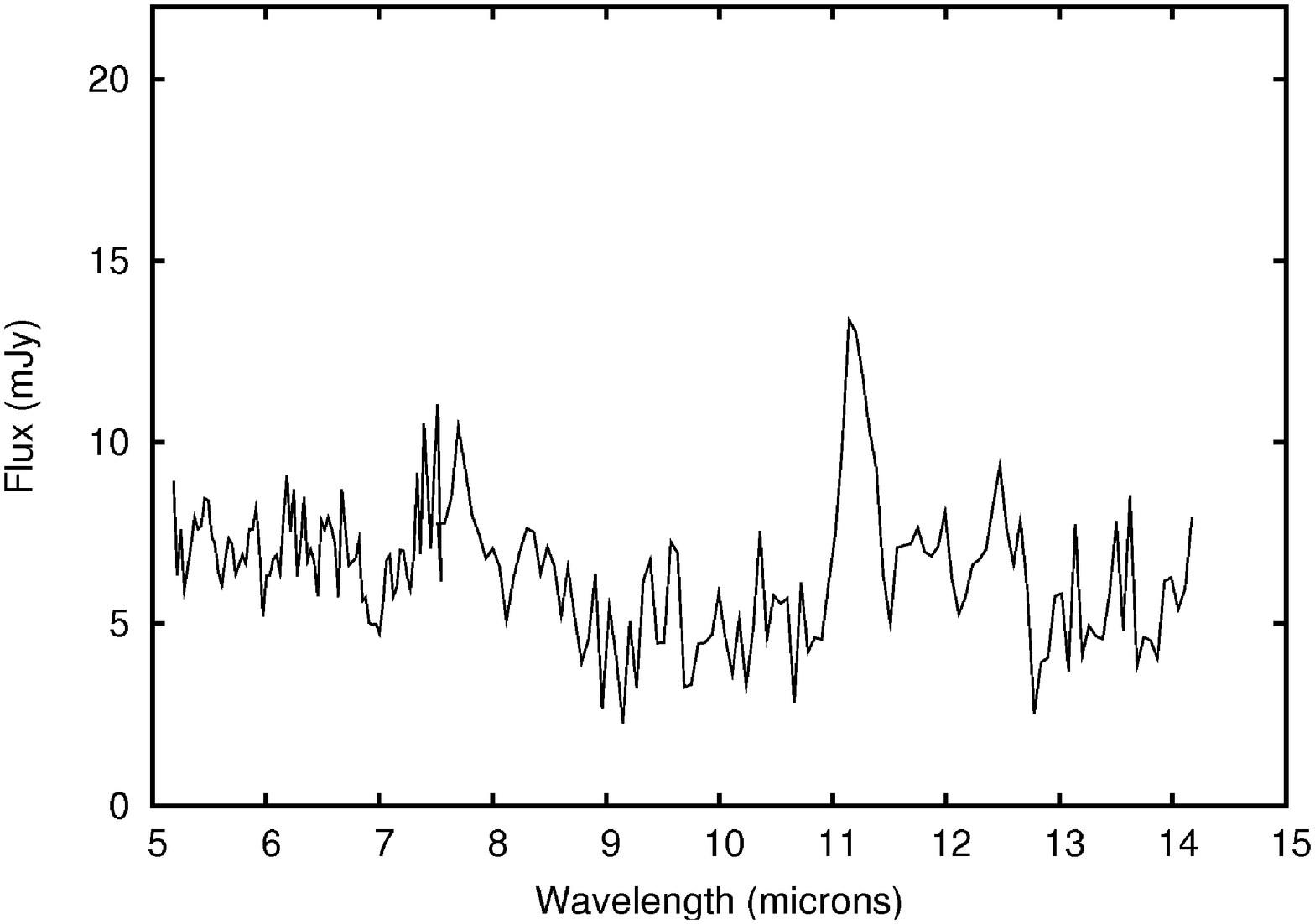}\\
\end{tabular}
\caption{\small Top: dereddened \grs1915\ radio/MIR/X-ray SEDs, built with the MJD~53511, MJD~53280, and MJD~53851 \rxte/PCA+HEXTE and IRS data (as well as archival VLA data in the case of MJD~53511), and fitted with the model \textit{phabs(diskir+gaussian+bbodyrad+bbodyrad+powerlaw)}.
\newline
Bottom: dereddened 5.20 to 14.50~\mic\ IRS spectra of \grs1915, obtained on MJD~ 53511, MJD~53280, and MJD~53851, used to fit the radio/MIR/X-ray SEDs. Note that from the left to the right, both SEDs and spectra are displayed by decreasing levels of irradiation.
}
\label{seds1915}
\end{center}
\end{figure*}

All the best-fit parameters are listed in Table~\ref{par1915spec}, and the fitted SEDs are displayed in Fig.~\ref{seds1915}. The lack of optical and NIR fluxes being a caveat to constrain irradiation, we added in each SED the \textit{K}-band magnitudes of \grs1915\ found in \citet{2007Neil}, the uncertainties corresponding to the magnitudes the day before and after our observations. These data were not used to fit the SEDs, but they show the emission level given in the \textit{K}-band by the irradiation component is not inconsistent with the measurements at the same epoch.

\section{Discussion}

\subsection{Origin of the dust component}

Our results suggest the existence of a cold dust component ($T_{\rm dust}\approx 300-500$~K) in the vicinity of \grs1915\ that likely interacts with the high-energy emission of the black hole binary. Moreover, the dust's temperature and radius appear to be roughly constant whatever the level of the X-ray emission, which implies that the dust is heated by the K giant companion star rather than the high-energy photons. And lastly, the average dust extension derived from the fits, $R_{\rm dust}\,\approx\,500$~\rsun\ or $3.5\times10^{13}$~cm, is roughly 5 times larger than the orbital separation and 10 times larger than the outer radius of the accretion disc, which means that the MIR emission due to the dust is produced well beyond the binary orbit and that the dust enshrouds the whole system.

Many red giant stars are known to be embedded in a dusty shell that originates from the slow and dense stellar winds \citep[see \textit{e.g.}][]{1978Hagen, 1986Zuckerman, 1987Morris, 2005Vanloon}, and this might be the case of the companion star of \grs1915. To check the consistency of the dust's temperature derived from the fits with the dusty stellar winds hypothesis, we can use the simple relation giving the expected temperature of a spherical dust shell in a 
thermodynamic equilibrium with a central star. Following \citet{2009Rahoui}, this temperature is:
\begin{equation}
  T_{\rm dust}=\left[\left (\frac{\pi^4}{60 Q_{\rm
        0}}\right )\left(\frac{h}{k}\right)^n\frac{1}{\Gamma(4+n)\zeta(4+n)}\left(\frac{R_{\rm \ast}}{R_{\rm
    dust}}\right)^2 T_{\rm \ast}^4\right]^\frac{1}{4+n}
\end{equation}
where $h$ and $k$ are the Planck and the Boltzmann constants, $\Gamma$ the gamma and $\zeta$ the Riemann zeta
functions, and $Q_0$ and $n$ such as the chromatic grain emissivity $Q_\nu$ is defined as $Q_\nu\,=\,Q_0 \nu^n$. Typical values for carboneous dust are $n\,=\,1.2$ and $Q_0\,=\,1.52\times10^{-8}r_{\rm g}$, where $r_{\rm g}$ is the 
average dust grain's radius \citep[see \textit{e.g.}][]{1984Draine,1998Robberto}. For 0.01~\mic$\,\le r_{\rm g} \le\,$0.1~\mic, as observed for interstellar dust grains \citep{1984Draine}, $T_\ast\,=\,4520$~K, $R_\ast\,=\,21$~\rsun, and $R_{\rm dust}\,\approx\,500$~\rsun, the inferred dust temperatures is about 315~K$\,\le T_{\rm dust} \le\,$490~K, which is consistent with our fits.  
\newline
 
However, the dust could also originate from a dusty disc-like circumstellar component. Such discs have already been invoked around 
some isolated first-ascent red giant stars, and a possible explanation for their presence could 
be the engulfment of an hypothetic low-mass companion when the star entered into the red giant phase 
\citep[see \textit{e.g.}][]{2003Jura, 2006Jura, 2009Melis}. They also 
have been detected around cataclysmic variables \citep{2004Dubus, 2006Howell, 2007Brinkworth, 2009Hoard}, 
and \citet{2006Muno} suggested their presence around A0620$-$00 and XTE~J1118$+$480 to explain the 8.00~\mic\ 
MIR excess in the emission of both sources while in quiescence. The two most common accepted explanations for the  
presence of circumbinary discs (CBDs) are (1) the dust was ejected from the binary with 
angular momentum during the common envelope phase, (2) it comes from a supernova fallback, as 
it was recently argued for the anomalous X-ray pulsar 4U~0142$+$61 \citep{2006Wang}. 

In the case of a flat and optically thick CBD irradiated by the companion star, the expected temperature at a radius $r$ is \citep{1997Chiang}:
\begin{equation}
T_{\rm CBD}(r)\,\approx\,\left (\frac{2}{3\pi} \right)^{\frac{1}{4}} \left (\frac{R_\ast}{r} \right )^{\frac{3}{4}}\,T_\ast
\end{equation}
For a CBD whose inner parts are truncated by tidal forces, the expected minimum inner radius is about $r_{\rm in}\,\approx\,1.7a$, 
where $a$ is the binary separation \citep{1994Artymowicz}. In the case 
of \grs1915, this leads to a maximum inner radius temperature of an hypothetical flat and optically thick CBD 
of $T_{\rm CBD}(r_{\rm in})\,\approx\,620~K$, which again is consistent with the MIR excess due to dust that we detected.

At this point, our data do not allow us to discriminate between a spherical and a disc geometry, but simple considerations 
confirm the presence of dust heated by the companion star and enshrouding the black hole binary. Moreover, it is worth noting that 
the MIR spectra of \grs1915\ are strongly similar in shape to those of some pre-main sequence or Herbig Ae/Be stars, silicate absorption and PAH features included 
\citep[see \textit{e.g.}][for such spectra]{2004Vanboekel, 2005Sloan, 2009Boersma, 2009Berne}. Such stars are known to exhibit a MIR excess due to the presence 
of an equatorial dusty disc within which PAH molecules are photoionised by the UV emission of the central star, and we suggest that this similarity strengthens 
the disc scenario for the dust distribution around \grs1915. 
\newline

Finally, the silicate absorption feature due to the diffuse interstellar medium is strongly correlated to the optical extinction as $A_{\rm V}\,=\,(18.5\pm2)\times \tau_{\rm 9.7}$ 
\citep{2003Draine}, where $\tau_{\rm 9.7}$ is the optical depth of the silicate absorption at 9.70~\mic. Following \citet{2007Chiar}, we took 
$\tau_{\rm 9.7}\,=\,-ln(F_{\rm 9.7}/F_{\rm continuum})$, where $F_{\rm 9.7}$ and $F_{\rm continuum}$ are 
the source and continuum's fluxes at 9.70~\mic, respectively.  The optical depth was computed on the spectrum obtained on MJD~53851, 
as this is the one for which X-ray reprocessing has the smallest contribution in the the MIR. The continuum was fitted using the ranges 
5.20$-$7.00~\mic\ and 13.00$-$14.50~\mic\ (in order to exclude the contribution of the silicate absorption feature) with a second order polynomial.
This method is strongly uncertain, especially concerning the continuum fitting, and the result should therefore be considered 
with caution. We nevertheless infer $A_{\rm V}\,=\,20.04\pm4.13$, which is consistent with the value found by \citet{2004Chapuis} for the optical extinction 
in the line of sight of \grs1915. We therefore conclude that the silicate absorption feature in the MIR spectrum of \grs1915\ is likely due to the diffuse interstellar medium.
 
\subsection{Importance of irradiation}

The results of the fitting of the three SEDs displayed in Fig.~\ref{seds1915} suggest that not only X-ray irradiation of the 
disc dominates the UV to NIR emission of \grs1915, but that it also extends to the MIR, where it is overcome by the dust component 
between about 6 to 10~\mic, depending on the accretion disc's flux. In the thermal state (MJD~53511), the MIR continuum is even strongly dominated 
by the thermalisation in the outer region of the soft X-ray emission from the inner parts. These results might once again emphasise the peculiarity 
of \grs1915. Indeed, only \citet{2007Migliari} previously proposed X-ray irradiation of the disc to explain the MIR emission of GRO~1655$-$40 
in the thermal state, and \citet{1994Paradijsa} excluded it for GRO~J0422+32, arguing that the accretion 
disc was not large enough to have outer regions sufficiently cold to emit at these wavelengths. This is not the case 
of \grs1915\ as the latter exhibits a large accretion disc, with an assessed outer radius of about $2.86\times 10^{12}$~cm. A strong contribution of 
the X-ray irradiation extending to the infrared was then expected, and \citet{2002Ueda} already argued that it was responsible for about 20$-$30\% 
of the $K-$band flux in the hard state, even in presence of compact jets.
\newline

Disc illumination therefore provides a consistent explanation for the variations of the MIR continuum of \grs1915. 
A caveat nevertheless forbids definitive conclusions. Indeed, X-ray reprocessing 
is thought to be the dominant contribution to the UV and optical fluxes of X-ray binaries, and any model needs these information to be well constrained. Their lack 
in the set of data we used to fit the SEDs is then a strong limitation to our interpretation, as the process could have artificially increased the emission 
at these wavelengths to better fit the MIR data. So, even if the fluxes given in \citet{2007Neil} are consistent with the ones derived from  our model, 
only the information on at least the quasi-simultaneous \textit{J}$-$, \textit{H}$-$, and \textit{K}$-$ bands magnitudes could strengthen our conclusions.

\section{Conclusion}

We presented a multi-wavelength study of \grs1915\ whose outcomes suggest that, in the absence of discrete or continuous ejecta, the 
MIR continuum of the source is mainly due to the X-ray irradiation of the accretion disc and to a photoionised dust component. This might 
have consequences on the interpretation of the MIR emission of microquasars in presence of compact jets. Indeed, dust might be 
ubiquitous around isolated compact objects and X-ray binaries, because of mass transfer during the common envelope phase or 
material from supernova fallback. If so, compact jets could contribute less than expected at infrared wavelengths, with perhaps a cutoff frequency 
in the millimeter domain. To confirm our results, it is therefore crucial to increase the sample of microquasars and systematically 
observe them through MIR spectroscopy, as this is the only way to obtain firm information on both their environment and their continuum. 
In particular, studying microquasars whose variation time scales are longer than the \grs1915\ ones, and which do not exhibit such rapid 
transitions between spectral states would strongly facilitate the multi-wavelength observations and would allow to reach definitive 
conclusions.

\acknowledgements
JR acknowledges partial funding by the European Commision under contract ITN 215212/ Black Hole Universe. 
This work was supported by the Centre National d'Etudes Spatiales (CNES), 
based on observations obtained through MINE: the Multi-wavelength INTEGRAL NEtwork.
This research has made use of NASA's Astrophysics
Data System, of the SIMBAD and VizieR databases operated at
CDS, Strasbourg, France, of products from the US Naval
Observatory catalogues, of products from the Two
Micron All Sky Survey as well as products from the Galactic Legacy Infrared Mid-Plane Survey
Extraordinaire, which is a \textit{Spitzer Space Telescope} Legacy Science Program.

\bibliography{./mybib}{}

\begin{deluxetable}{ccccccc}
\tabletypesize{\scriptsize}
\rotate
\tablecaption{\small Summary of all the \spitzer's data of \grs1915~we made use of in this study\label{logobs}}
\tablewidth{0pt}
\tablehead{\colhead{IRAC}&\colhead{IRS}&\colhead{\rxte}&\colhead{\intl}&\colhead{Ryle}&\colhead{Ryle fluxes (mJy)}&\colhead{Class}}
\startdata
\nodata&53280.275&53280.241$-$53280.348&\nodata&53280.636$-$53280.746&$<$1&$\rho$\\
53284.176&\nodata&53284.177$-$53284.236&\nodata&53283.673$-$53283.706&$<$1&$\rho$\\
\nodata&53299.226&\nodata&\nodata&53298.587$-$53298.663&$<$1&\nodata\\
53308.741&\nodata&\nodata&\nodata&53308.811$-$53308.841&$<$1&\nodata\\
\nodata&53484.099&\nodata&\nodata&53484.027$-$53484.169&76$-$91&\nodata\\
53496.287&\nodata&\nodata&\nodata&53496.209$-$53496.230&37$-$43&\nodata\\
53500.571&\nodata&53500.326$-$53500.430&\nodata&53500.202$-$53500.358&22$-$32&$\delta$\\
\nodata&53511.697&\nodata&\nodata&\nodata&\nodata&\nodata\\
53636.571&\nodata&\nodata&\nodata&\nodata&\nodata&\nodata\\
\nodata&53660.072&53659.982$-$53660.079&\nodata&53659.577$-$53659.906&27$-$47&$\phi$\\
\nodata&53661.182&53661.706$-$53661.856&\nodata&53660.602$-$53660.881&18$-$55&$\phi$\\
53676.070&\nodata&\nodata&53676.247$-$53677.488&53675.583$-$63675.641&64$-$75&$\chi$, $\mu$, $\beta$\\
\nodata&53851.419&53851.307$-$53851.319&\nodata&53851.102$-$53851.320&$<$3&$\chi$\\
53857.587&\nodata&\nodata&\nodata&\nodata&\nodata&\nodata\\
\nodata&53874.818&\nodata&\nodata&\nodata&\nodata&\nodata\\
53890.827&\nodata&53892.047$-$53892.063&\nodata&53888.239$-$53888.261&$<$1&$\rho$\\
\enddata
\tablecomments{For each instrument, we give the day of observation (in MJD), and, when available, we give the day of quasi-simultaneous coverage with \rxte, \intl, and/or the Ryle telescope, as well as the Ryle flux level (15~GHz) in mJy. Moreover, when high-energy observations were available, we give the spectral class of the source as defined in \citet{2000Belloni}}. 
\end{deluxetable}

\begin{deluxetable}{cccccc}
\tabletypesize{\footnotesize}
\tablecaption{\small Absorbed fluxes (in mJy) of \grs1915\ in the four IRAC filters\label{fluxirac1915}}
\tablewidth{0pt}
\tablehead{\colhead{MJD}&\colhead{3.59~\mic}&\colhead{4.50~\mic}&\colhead{5.80~\mic}&\colhead{8.00~\mic}}
\startdata
53284&5.70$\pm$0.17&5.32$\pm$0.16&4.80$\pm$0.15&3.12$\pm$0.10\\
53308&5.26$\pm$0.16&4.85$\pm$0.15&4.32$\pm$0.13&2.95$\pm$0.10\\
53496&8.69$\pm$0.26&8.12$\pm$0.24&7.61$\pm$0.23&5.03$\pm$0.16\\
53500&10.28$\pm$0.31&10.09$\pm$0.30&8.80$\pm$0.27&6.14$\pm$0.19\\
53636&6.20$\pm$0.19&5.88$\pm$0.18&5.33$\pm$0.16&3.77$\pm$0.12\\
53676&10.70$\pm$0.32&10.31$\pm$0.31&9.15$\pm$0.28&6.41$\pm$0.20\\
53857&4.82$\pm$0.15&4.75$\pm$0.14&4.23$\pm$0.13&2.87$\pm$0.09\\
53890&5.16$\pm$0.16&4.92$\pm$0.15&4.36$\pm$0.13&3.05$\pm$0.10\\
\enddata
\tablecomments{Uncertainties are given at 1$\sigma$ and include 3\% systematic errors.}
\end{deluxetable}

\begin{deluxetable}{ccccccc}
  \tablewidth{0pt}
  \tabletypesize{\footnotesize}
  \rotate
  \tablecaption{\small Best parameters obtained from the fit of the MJD~53280, MJD~53284, MJD~53500, MJD~53676, MJD~53851, and MJD~53890 high-energy spectra of \grs1915\label{par1915he}}
  \tablehead{\colhead{Parameters}&\colhead{MJD~53280}&\colhead{MJD~53284}&\colhead{MJD~53500}&\colhead{MJD~53676}&\colhead{MJD~53851}&\colhead{MJD~53890}}
  \startdata
  $kT_{\rm disc}$(keV)&1.27$_{-0.03}^{+0.02}$&1.20$_{-0.02}^{+0.02}$&1.85$_{-0.02}^{+0.02}$&2.05$_{-0.04}^{+0.04}$&0.81$_{-0.05}^{+0.05}$&1.28$_{-0.09}^{+0.07}$\\
  $Norm$&$594.0_{-36.9}^{+40.7}$&$765.9_{-46.2}^{+51.8}$&$267.5_{-11.3}^{+12.0}$&$90.99_{-8.18}^{+9.82}$&$1193.0_{-264.6}^{+386.6}$&$396.8_{-53.3}^{+78.5}$\\
  $kT_{\rm e}$ (keV)&$158.8_{-3.6}^{+3.7}$&$155.2_{-2.9}^{+3.0}$&$118.1_{-7.9}^{+8.4}$&$158.2_{-7.3}^{+7.8}$&$27.3_{-6.6}^{+13.7}$&$131.7_{-9.9}^{+10.9}$\\
  $\tau$&0.01 (\textit{frozen})&0.01 (\textit{frozen})&0.01 (\textit{frozen})&0.01 (\textit{frozen})&0.95$_{-0.39}^{+0.37}$&0.01 (\textit{frozen})\\
  $F_{\rm total}$\tablenotemark{a} ($\times10^{-8}$~erg~cm$^{-2}$~s$^{-1}$)&3.08&3.49&4.64&4.59&1.71&2.13\\
  $F_{\rm disc}$\tablenotemark{a} ($\times10^{-8}$~erg~cm$^{-2}$~s$^{-1}$)&1.22&1.19&3.75&3.74&0.20&0.85\\
  $\chi^2$(d.o.f)&0.93 (53)&1.09 (53)&1.29 (53)&1.65 (49)&1.31 (53)&1.25 (53)\\
  \enddata
  \tablecomments{The best-fit model is \textit{phabs(diskbb+gaussian+comptt)}, and the errorbars are given at the 90\% confidence level. These spectra were built with \rxte/PCA+HEXTE data, except on MJD~53676 for which we used \intl/JEM$-$X+ISGRI data.}
  \tablenotetext{a}{$F_{\rm total}$ and $F_{\rm disc}$ are the total and disc unabsorbed fluxes, extrapolated to 3.0$-$200.0~keV.}
\end{deluxetable}

\begin{deluxetable}{ccc}
  \tablewidth{0pt}
  \tabletypesize{\footnotesize}
  \tablecaption{\small Best parameters from the fit of the MJD~53660 and MJD~53661 \rxte/PCA+HEXTE spectra of \grs1915\label{par1915he2}}
  \tablehead{\colhead{Parameters}&\colhead{MJD~53660}&\colhead{MJD~53661}}
  \startdata
  $kT_{\rm seed}$ (keV)&$0.53_{-0.04}^{+0.10}$&$0.72_{-0.04}^{+0.05}$\\
  $kT_{\rm e}$ (keV)&$6.17_{-0.94}^{+1.84}$&$5.48_{-0.56}^{+0.72}$\\
  $\tau$&$2.08_{-0.42}^{+0.44}$&$2.04_{-0.25}^{+0.24}$\\
  $\Gamma$\tablenotemark{a}&$2.71_{-0.87}^{+0.30}$&$2.87_{-0.15}^{+0.10}$\\
  $F_{\rm total}$\tablenotemark{b} ($\times10^{-8}$erg~cm$^{-2}$~s$^{-1}$)&3.08&4.43\\
  $\chi^2$ (d.o.f)&0.82 (53)&0.74 (53)\\
  \enddata
  \tablecomments{The best-fit model is \textit{phabs(comptt+gaussian+powerlaw)}, and the errorbars are given at the 90\% confidence level.}
  \tablenotetext{a}{$\Gamma$ is the power law photon index.}
  \tablenotetext{b}{$F_{\rm total}$ is the total unabsorbed flux, extrapolated to 3.0$-$200.0~keV.}
\end{deluxetable}

\begin{deluxetable}{ccccccc}
\tabletypesize{\footnotesize}
\tablecaption{\small List of all detected features for each IRS spectrum of \grs1915 \label{raie1915}}
\tablewidth{0pt}
\tablehead{\colhead{Features}&\colhead{$\lambda$}&\colhead{$\lambda_{\rm fit}$}&\colhead{$\mathring{W}$}&\colhead{\textit{FWHM}}&\colhead{Flux}&\colhead{\textit{SNR}}\\
\nodata&\colhead{(\mic)}&\colhead{(\mic)}&\colhead{(\mic)}&\colhead{(\mic)}&\colhead{$\times 10^{-21}$~W~cm$^{-2}$}&\nodata}
\startdata
&&&\multicolumn{1}{c}{MJD~53280}\\
\hline
\ion{H}{1}&7.508&7.522$\pm$0.004&-0.041&0.113$\pm$0.005&1.030$\pm$0.119&11.29\\
PAH&7.800&7.801$\pm$0.008&-0.032&0.228$\pm$0.009&1.266$\pm$0.137&15.93\\
PAH&11.250&11.251$\pm$0.004&-0.193&0.178$\pm$0.004&0.622$\pm$0.041&18.04\\
\ion{He}{2}~?&12.367&12.353$\pm$0.003&-0.090&0.163$\pm$0.003&0.424$\pm$0.026&27.54\\
\ion{H}{1}~?&12.587&12.590$\pm$0.011&-0.075&0.239$\pm$0.014&0.576$\pm$0.089&14.65\\
\hline
&&&\multicolumn{1}{c}{MJD~53299}\\
\hline
\ion{H}{1}&7.508&7.519$\pm$0.003&-0.058&0.134$\pm$0.004&1.546$\pm$0.169&34.30\\
PAH&7.600&7.621$\pm$0.002&-0.034&0.100$\pm$0.003&1.432$\pm$0.078&25.37\\
PAH&11.250&11.206$\pm$0.003&-0.227&0.189$\pm$0.003&0.599$\pm$0.039&36.64\\
\hline
&&&\multicolumn{1}{c}{MJD~53484}\\
\hline
\ion{H}{2}&7.460&7.458$\pm$0.004&-0.050&0.122$\pm$0.007&1.546$\pm$0.06&68.28\\
PAH&7.700&7.722$\pm$0.008&-0.097&0.240$\pm$0.009&2.581$\pm$0.190&16.55\\
PAH&11.250&11.299$\pm$0.014&-0.110&0.143$\pm$0.012&0.463$\pm$0.060&8.85\\
\ion{H}{1}&12.587&12.599$\pm$0.007&-0.140&0.194$\pm$0.007&0.670$\pm$0.076&12.49\\
\hline
&&&\multicolumn{1}{c}{MJD~53511}\\
\hline
\ion{H}{2}&7.460&7.460$\pm$0.002&-0.029&0.087$\pm$0.002&0.903$\pm$0.073&17.47\\
$[$\ion{Ne}{2}$]$&12.813&12.803$\pm$0.001&-0.080&0.136$\pm$0.001&0.427$\pm$0.011&60.68\\
\hline
&&&\multicolumn{1}{c}{MJD~53660}\\
\hline
PAH&7.700&7.719$\pm$0.021&-0.029&0.264$\pm$0.017&1.293$\pm$0.138&16.11\\
PAH&11.250&11.279$\pm$0.011&-0.077&0.256$\pm$0.005&0.692$\pm$0.035&46.56\\
\hline
&&&\multicolumn{1}{c}{MJD~53661}\\
\hline
PAH&7.800&7.812$\pm$0.005&-0.051&0.187$\pm$0.006&1.389$\pm$0.111&14.66\\
PAH&11.250&11.311$\pm$0.005&-0.100&0.208$\pm$0.005&0.644$\pm$0.050&17.89\\
\ion{H}{1}&12.157&12.136$\pm$0.004&-0.094&0.183$\pm$0.006&0.625$\pm$0.055&17.69\\
\ion{H}{1}&12.611&12.606$\pm$0.011&-0.089&0.223$\pm$0.012&0.667$\pm$0.099&11.21\\
\hline
&&&\multicolumn{1}{c}{MJD~53851}\\
\hline
PAH&7.800&7.773$\pm$0.006&-0.043&0.119$\pm$0.006&0.853$\pm$0.126&11.29\\
PAH&11.250&11.269$\pm$0.009&-0.347&0.281$\pm$0.010&1.188$\pm$0.120&11.42\\
\hline
&&&\multicolumn{1}{c}{MJD~53874}\\
\hline
PAH&7.800&7.831$\pm$0.010&-0.026&0.121$\pm$0.010&1.152$\pm$0.112&11.76\\
PAH&11.250&11.247$\pm$0.001&-0.076&0.115$\pm$0.002&0.662$\pm$0.022&45.99\\
\ion{H}{1}&12.587&12.568$\pm$0.011&-0.101&0.253$\pm$0.028&0.670$\pm$0.128&11.02\\
\enddata
\tablecomments{We give the name of the feature, its laboratory ($\lambda$) and measured ($\lambda_{\rm fit}$) wavelengths, 
its equivalent width $\mathring{W}$, its full-width at half-length ($FWHM$), its flux, and its signal-to-noise ratio ($SNR$).}
\end{deluxetable}

\begin{deluxetable}{cccc}
\tablewidth{0pt}
\tabletypesize{\footnotesize}
\tablecaption{\small  Best parameters obtained from the fits of the \grs1915\ MIR to X-ray dereddened SEDs\label{par1915spec}}
\footnotesize
\tablehead{\colhead{Parameters}&\colhead{MJD~53280}&\colhead{MJD~53511}&\colhead{MJD~53851}}
\startdata
$kT_{\rm disc}$ (keV)&1.33$_{-0.02}^{+0.02}$&1.88$_{-0.01}^{+0.03}$&0.75$_{-0.06}^{+0.05}$\\
$Norm$&$520.08_{-30.24}^{+32.59}$&$251.46_{-6.14}^{+6.36}$&$1756.80_{-429.51}^{+677.31}$\\
$lrout$ ($R_{\rm in}$)&6.01&6.17&5.74\\
$\Gamma$&$2.74_{-0.03}^{+0.02}$&$2.96_{-0.07}^{+0.07}$&$2.13_{-0.03}^{+0.02}$\\
$Lc/Ld$&$0.46_{-0.01}^{+0.01}$&$0.067_{-0.001}^{+0.002}$&$1.39_{-0.05}^{+0.06}$\\
$fin$&0.1 (\textit{frozen})&0.3 (\textit{frozen})&0.1 (\textit{frozen})\\
$fout$ ($10^{-4}$)&$29.39_{-9.71}^{+11.84}$&$368.53_{-89.20}^{+105.58}$&$7.34_{-5.42}^{+7.46}$\\
$T_{\rm dust}$ (K)&$396.55_{-36.65}^{+44.44}$&$306.66_{-42.24}^{+87.44}$&$461.16_{-60.35}^{+86.50}$\\
$R_{\rm dust}$ ($R_{\odot}$)&$538.40_{-91.32}^{+95.26}$&$743.66_{-178.52}^{+175.51}$&$375.17_{-72.91}^{+87.91}$\\
$\Gamma_{\rm radio}$&\nodata&$-0.85^{+0.05}_{-0.09}$&\nodata\\
$\chi^2$ (d.o.f)&1.28 (187)&1.12 (190)&1.64 (187)\\
\enddata
\tablecomments{The best-fit model is \textit{phabs(diskir+gaussian+bbodyrad+bbodyrad+powerlaw)}, and the errorbars are given at the 90\% confidence level. The SEDs were built with the MJD~53280, MJD~53511, and MJD~53851 \rxte/PCA+HEXTE and IRS data, as well as archival VLA data in the case of MJD~53511.}
\end{deluxetable}

\end{document}